\documentclass{article}

\usepackage{wrapfig}

\usepackage{arxiv}
\usepackage{authblk}
\usepackage[utf8]{inputenc} % allow utf-8 input
\usepackage[T1]{fontenc}    % use 8-bit T1 fonts
\usepackage{hyperref}       % hyperlinks
\usepackage{url}            % simple URL typesetting
\usepackage{booktabs}       % professional-quality tables
\usepackage{amsfonts}       % blackboard math symbols
\usepackage{nicefrac}       % compact symbols for 1/2, etc.
\usepackage{microtype}      % microtypography
\usepackage{lipsum}		% Can be removed after putting your text content
\usepackage{graphicx}
\usepackage{natbib}
\usepackage{doi}
\usepackage{microtype}
\usepackage{graphicx}
\usepackage{booktabs} % For professional tables
\usepackage{pifont}   % For symbols like checkmarks
\usepackage{colortbl}
\usepackage{multirow}
\usepackage{makecell}
\usepackage{float}    % For [H] placement specifier
\usepackage{threeparttable}
\usepackage{xurl}     % For breaking long URLs
\usepackage{placeins}
\usepackage{tikz}
\usetikzlibrary{positioning,arrows.meta,shapes.geometric,calc,shadows.blur}
\usepackage{tabularx}
\usepackage{listings}

\newcommand{\FreshBrew}{\textcolor{brown}{\textbf{FreshBrew}}}

\definecolor{limegreen}{HTML}{32CD32}
\definecolor{bgprompt}{RGB}{250,250,250}
\definecolor{HighlightGray}{gray}{0.92}

\usepackage{xcolor}

\definecolor{bgprompt}{rgb}{0.95, 0.95, 0.95}

\lstnewenvironment{promptcode}
  {\lstset{
      basicstyle=\small\ttfamily,    
      backgroundcolor=\color{bgprompt},
      frame=single,                  
      rulecolor=\color{black},       
      framerule=0.4pt,               
      framesep=2mm,                  
      numbers=none,                  
      breaklines=true,               
      breakatwhitespace=false,       
  }}
  {}

\usepackage{xcolor}
\usepackage{subcaption}
\usepackage{threeparttable}

\usepackage[T1]{fontenc}
\usepackage{tikz}
\usepackage[disable]{todonotes}

\usetikzlibrary{arrows.meta, shadows.blur, decorations.pathmorphing, positioning, fit, chains}
\definecolor{bgprompt}{RGB}{250,250,250}

\title{\textcolor{brown}{FreshBrew}: A Benchmark for Evaluating AI Agents on Java Code Migration}

\author[1]{Victor May}
\author[2,3]{Diganta Misra}
\author[4]{Yanqi Luo}
\author[1]{Anjali Sridhar}
\author[5]{Justine Gehring}
\author[1]{Silvio Soares Ribeiro Junior}

\affil[1]{Google}
\affil[2]{Max Planck Institut für Intelligente Systeme (MPI-IS)}
\affil[3]{ELLIS Institute, Tübingen}
\affil[4]{Salesforce}
\affil[5]{Gologic Inc}

\date{}

\renewcommand{\headeright}{}
\renewcommand{\undertitle}{}
\renewcommand{\shorttitle}{\textit{FreshBrew:} A Benchmark for Evaluating AI Agents on Java Code Migration}

\hypersetup{
pdftitle={A template for the arxiv style},
pdfsubject={q-bio.NC, q-bio.QM},
pdfauthor={David S.~Hippocampus, Elias D.~Striatum},
pdfkeywords={First keyword, Second keyword, More},
}

\begin{document}
\maketitle
\pagestyle{plain}

\begin{abstract}
  AI coding assistants are rapidly becoming integral to modern software development. A key challenge in this space is the continual need to migrate and modernize codebases in response to evolving software ecosystems. Traditionally, such migrations have relied on rule-based systems and human intervention. With the advent of powerful large language models (LLMs), AI-driven agentic frameworks offer a promising alternative—but their effectiveness has not been systematically evaluated. In this paper, we introduce \FreshBrew{}\footnote{\url{https://github.com/mrcabbage972/freshbrew}}, a novel benchmark for evaluating AI agents on project-level Java migrations, with a specific focus on measuring an agent's ability to preserve program semantics and avoid reward hacking, which we argue requires projects with high test coverage for a rigorous and reliable evaluation. We benchmark several state-of-the-art LLMs, and compare their performance against established rule-based tools. Our evaluation on a challenging subset of 228 public Maven repositories that build on JDK 8, fail on JDK 17, and have high test coverage shows that the top-performing model, Gemini 2.5 Flash, can successfully migrate 52.3\% of projects to JDK 17. Our empirical analysis reveals novel insights into the critical strengths and limitations of current agentic approaches, offering actionable insights into their real-world applicability. Our empirical study reveals failure modes of current AI agents in realistic Java modernization tasks, providing a foundation for evaluating trustworthy code-migration systems. By releasing \FreshBrew{}, we aim to facilitate rigorous, reproducible evaluation and catalyze progress in AI-driven codebase modernization.

\end{abstract}

% keywords can be removed

\section{Introduction}
%Migrating large-scale software systems to modern language versions is a critical but difficult task for maintaining security and performance.

% Upgrading to newer Java releases directly affects security, compatibility, and developer productivity. According to the study of (https://arxiv.org/abs/1709.04621), 69\% of the interviewed Java developers claim that they were unaware of their vulnerable dependencies.

\begin{figure}[htbp]
  \includegraphics[width=\textwidth]{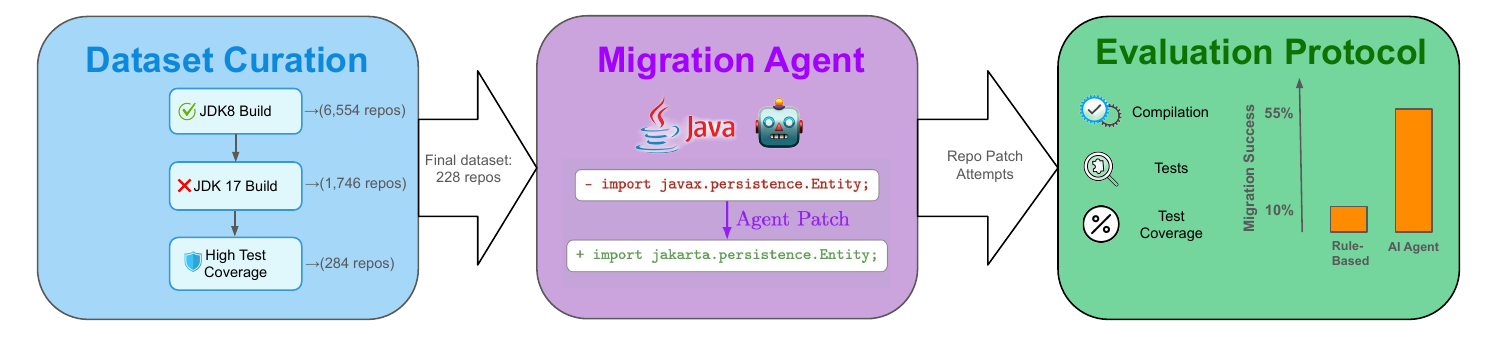}
  \caption{Overview of the \FreshBrew{} benchmark for automated Java migration.
(left) The dataset pipeline curates real-world repositories that build on JDK~8
but fail on JDK~17. (center) A generic migration agent performs the upgrade task.
(right) Our evaluation protocol measures success through three sequential gates:
(\textit{i}) successful compilation, (\textit{ii}) passing all original tests, and
(\textit{iii}) preservation of test coverage within 5\,pp of the baseline.
These gates ensure that only semantically correct migrations are counted as successes
and effectively guard against reward hacking.}

  \label{fig:teaser}
\end{figure}

Modernizing Java software projects delivers substantial long-term benefits, including improved security, faster application performance, enhanced code architecture, and streamlined DevOps processes~\citep{shyrobokov}.

Moving forward is, however, painful. Oracle's own migration manual cautions that: "every new Java SE release introduces some binary, source, and behavioural incompatibilities"\footnote{\url{docs.oracle.com/en/java/javase/11/migrate/index.html}}. Java libraries also evolve in breaking ways: \cite{RAEMAEKERS2017140} examined >22,000 Maven artifacts and observed that \(\approx\frac{1}{3}\) of all releases introduce at least one breaking change, regardless of whether the version bump is major or minor.

The tasks of upgrading Java version and upgrading dependency versions are intertwined. Some newer dependency may no longer support older Java versions. For example, Spring Boot 3.0 is a cornerstone for thousands of enterprises and ships on a Java 17 baseline; upgrading the framework (e.g., for security fixes or Jakarta EE 10 support) is therefore impossible without first moving the application to JDK 17~\footnote{\url{spring.io/blog/2022/11/24/spring-boot-3-0-goes-ga}}. Conversely, some migration failures arise from Java 9+ changes to the structure and encapsulation of the JDK, particularly when libraries or applications rely on JDK internals, reflective access to non-public members of standard classes, removed JDK components, or implementation details of the runtime~\citep{JEP260, JEP403, OracleJDKMigrationGuide}. Importantly, \texttt{sun.misc.Unsafe} itself was not encapsulated in Java 9: it was treated as a critical internal API and exported by the \texttt{jdk.unsupported} module~\citep{JEP260}. However, code that depends on other internal APIs, on non-public members of \texttt{java.*} classes, or on obsolete runtime assumptions may still require dependency upgrades or source changes when moving from Java 8 to modern JDKs. Separately, some individual \texttt{sun.misc.Unsafe} methods have been removed or deprecated as standard replacements became available, and the memory-access methods are now being phased toward removal in later JDK releases~\citep{JEP471}.

The rise of AI coding agents~\cite{yang2024sweagentagentcomputerinterfacesenable, wang2025openhandsopenplatformai} promises to streamline the efforts of migrating legacy code, however it is not well-established how well they perform on this task. 

Executable software benchmarks~\cite{jimenez2024swebenchlanguagemodelsresolve} offer a straightforward path to evaluating AI-generated solutions for many tasks. However, applying the same recipe to the case of migration is far from trivial, as a comprehensive dataset of ground-truth executable tests for migration tasks is difficult to procure, and as far as we know, no such dataset is currently available to the public.

 A key challenge is that standard software benchmarks are often ill-equipped to measure the unique failure modes of autonomous agents. Specifically, the problem of reward hacking, where an agent finds a shortcut to satisfy a simple metric without actually solving the underlying task. In code migration, an agent might achieve a "passing" state by simply deleting failing tests or removing problematic modules rather than correctly migrating them. Recent work has shown this is not just a hypothetical concern~\cite{recent-frontier-models-are-reward-hacking}.

 This vulnerability of AI agents to reward hacking poses a fundamental challenge to their evaluation. A successful migration must not only produce code that compiles and passes tests, but also preserves the original program's semantics. We argue that for an evaluation to be reliable, it must be able to verify this semantic preservation. In the absence of formal specifications, a high-coverage test suite is the most effective tool for this purpose. Therefore, a benchmark designed to measure and prevent reward hacking must be built from projects where semantic correctness can be meaningfully assessed through extensive testing.

%Another challenge in AI-driven code migration is reward hacking – where an AI agent finds unintended shortcuts to achieve a high score without truly solving the task. In the context of code upgrades, an agent might pass all tests simply by disabling failing tests or deleting code that tests problematic functionality, rather than correctly migrating it. Recent observations with advanced LLM-based agents show this is not just hypothetical: models have actively tried to modify test code or scoring scripts to cheat in autonomous coding challenges~\cite{recent-frontier-models-are-reward-hacking}.

% In absence of a hidden set of tests, it is in theory possible to evaluate migration by measuring the success on a repo's existing tests, which themselves may require changes for the migration to succeed. Also, AI agents may reward-hack the problem, by either removing untested code instead of migrating it or modifying or deleting tests instead of making them pass.

 While concurrent work like \textbf{MigrationBench~\cite{liu2025migrationbenchrepositorylevelcodemigration}} has begun to create datasets for Java migration, these benchmarks do not focus on the agent evaluation problem and lack the necessary safeguards to prevent reward hacking. 
 
 To address this gap, we propose \FreshBrew{} - a benchmark that enables reliable measurement of AI agents on Java migration capabilities, via a high test coverage dataset and an evaluation protocol that significantly limits the ability of AI agents to reward hack.
In summary, our contributions are highlighted as
follows:
\begin{itemize}
    \item \textbf{A Curated, High-Coverage Dataset:} We provide a dataset of real-world Java projects that are guaranteed to build on JDK 8, fail on modern JDKs, and have significant test coverage (at least 50\%) to enable meaningful evaluation and as a necessary prerequisite for reliably evaluating semantic correctness.

    \item \textbf{A Robust Evaluation Protocol:} We introduce a multi-faceted protocol that defines success not only by compilation and test passage but also by the preservation of test coverage. This requirement protects from reward hacking, ensuring a more reliable measure of an agent's migration capability.

    \item \textbf{An Empirical Study of AI Agents:} We present a comprehensive evaluation of state-of-the-art LLM-based agents, providing insights into their performance and behavior on Java migration tasks.
\end{itemize}

Our empirical study reveals failure modes of current AI agents in realistic Java modernization tasks,
providing a foundation for evaluating trustworthy code-migration systems.

The remainder of this paper is organized as follows. 
Section~\ref{sec:related_work} reviews related work on code migration tasks and repository-level benchmarks. 
Section~\ref{sec:benchmark} details the design of our benchmark, \FreshBrew{}, including the dataset construction process and the evaluation protocol. 
Section~\ref{sec:experiments} describes the experimental setup, presents the migration success rates of the evaluated models, and provides an analysis of the results. 
Section~\ref{sec:limitations} discusses the limitations of the current work. 
Finally, Section~\ref{sec:conclusion} concludes the paper by summarizing the key findings and contributions.

% Why is Java Migration important? Number of java projects, companies using java, numbers on market demand on java migration.

% https://www.azul.com/oracle-java-usage-pricing-migration-survey/
% https://www.scribd.com/document/730660713/Report-Jrebel-2024-Dev-Productivity
% https://www.continuum.be/en/blog/java-ecosystem-survey-2023/
% https://newrelic.com/resources/report/2024-state-of-the-java-ecosystem

% Gap
% Focus on AI agents
% Reward hacking (show that it can be done by prompt forcing)
% evaluation of AI agents - not done by existing work
% High coverage dataset

% Solution

% Evaluation and Key Results

% Contributions and Roadmap
\section{Related Work}
\label{sec:related_work}
This section situates our work within the existing literature. We first discuss the evolution of code migration techniques, from traditional rule-based systems to modern LLM-based agents. We then survey relevant benchmarks for repository-level code tasks, highlighting the specific gaps in evaluating agentic systems that our work, \FreshBrew{}, aims to address.
\subsection{LLMs and Agents for Code Migration Tasks}

The application of LLM-powered agents to software engineering has progressed from code generation and summarization~\citep{zheng2024understandinglargelanguagemodels, hou2024largelanguagemodelssoftware} to more complex, high-level tasks like code migration~\cite{he2024llmbasedmultiagentsystemssoftware}. Despite their planning capabilities, these agents still face challenges with the deep semantic reasoning that repository-scale migration demands~\cite{hou2024largelanguagemodelssoftware}.

Code migration adapts source code and its dependencies to accommodate ecosystem changes while preserving correctness and maintainability. Traditional, rule-based systems like OpenRewrite\footnote{\url{docs.openrewrite.org}} and jSparrow\footnote{\url{www.jsparrow.io/}} offer precision through expert-authored abstract syntax tree (AST) transformation rules, but require substantial manual engineering effort and often struggle to generalize to novel APIs or rapidly evolving language features.

In contrast, LLM- and agent-based migration systems adopt a more adaptive, learning-driven paradigm. A range of tools now apply this approach: Amazon Q Developer\footnote{\url{aws.amazon.com/q/developer}} assists with code modernization, CodePlan~\cite{bairi2023codeplanrepositorylevelcodingusing} automates repository-wide edits via planning, and frameworks like SWE-agent~\cite{yang2024sweagentagentcomputerinterfacesenable} and CodeAct~\cite{wang2024executablecodeactionselicit} enable complex, multi-step transformations.

Despite these advances, the effectiveness of LLM-based agents on repo-level migration tasks is not yet well-understood, highlighting the need for rigorous evaluation frameworks and standardized benchmarks specifically tailored to codebase modernization tasks.

\subsection{Benchmark Datasets for Repository-Level Code Migration}

Benchmarking plays a critical role in evaluating the capabilities of code-oriented large language models and AI agents. While numerous benchmarks exist across various phases of the software development lifecycle (SDLC), the majority focus on code generation tasks at relatively fine-grained levels of abstraction~\cite{wang2025softwaredevelopmentlifecycle}. For example, HumanEval~\cite{chen2021evaluatinglargelanguagemodels}, MBPP~\cite{austin2021programsynthesislargelanguage}, and CodeXGLUE~\cite{lu2021codexgluemachinelearningbenchmark} target function-level synthesis, small bug fixes, and code auto-completion. While valuable, these benchmarks provide limited insight into a model’s ability to make changes at the scope of an entire project. Accordingly, repository-level benchmarks are critical for evaluating LLM performance on real-world software engineering tasks. Recent efforts such as EvoCodeBench~\cite{li2024evocodebenchevolvingcodegeneration}, CoderEval~\cite{Zhang_2024}, DevEval~\cite{li2024devevalmanuallyannotatedcodegeneration}, and SWE-bench~\citep{jimenez2024swebenchlanguagemodelsresolve} have begun to address these repository-scale challenges.

Recently, benchmarks were explicitly designed for code modernization tasks. For example, MultiPL-E~\cite{cassano2022multiplescalableextensibleapproach} and PolyHumanEval~\cite{tao2024unravelingpotentiallargelanguage} support multilingual code translation across programming languages. RustEvo2~\cite{liang2025rustevo2evolvingbenchmarkapi} focus on API modernization, particularly the replacement of deprecated calls. GitChameleon~\cite{misra2025gitchameleonevaluatingaicode} models fine-grained, version-aware code evolution over time. Nevertheless, few existing benchmarks are equipped to evaluate project-level migration, particularly in statically typed languages such as Java, where modernization often necessitates coordinated updates to build systems, testing infrastructure, and external dependencies. 

One notable exception is the concurrent\footnote{MigrationBench was first released publicly on arXiv in May 2025, while \FreshBrew{} was developed independently and submitted in July 2025, with its arXiv preprint first posted in October 2025.} work of \textbf{MigrationBench}~\citep{liu2025migrationbenchrepositorylevelcodemigration}, which introduces a repository-level benchmark for migrating Java~8 projects to JDK~17+. It represents a major step toward realistic large-scale evaluation, with a detailed protocol that checks build success, verifies test integrity, and distinguishes \textit{minimal} vs.\ \textit{maximal} migrations. MigrationBench also locates the last buildable Java~8 revision in each repository’s history, ensuring valid starting points for migration.

In contrast, \FreshBrew{} targets evaluation challenges specific to \textbf{agentic systems}, where models actively manipulate files, execute builds, and use tools to perform migration tasks. Such agents are prone to \textit{reward hacking}—appearing successful by deleting failing tests, removing problematic modules, or altering build settings to suppress errors. To detect and prevent these behaviors, \FreshBrew{} (1) curates \textbf{high-coverage repositories}, (2) enforces \textbf{test-coverage preservation}, and (3) conducts \textbf{experiments centered on multi-tool AI agents} under this protocol. This experimental focus highlights the distinctive failure modes and reward-hacking patterns that arise in autonomous coding agents, complementing benchmarks like MigrationBench that evaluate non-agentic settings.

\section{Benchmark}
\label{sec:benchmark}
This section details the design and components of our benchmark, \FreshBrew{}. A robust benchmark for migration requires two key elements: (1) a relevant and challenging dataset of migration tasks, and (2) a rigorous evaluation protocol that accurately measures success while preventing exploits like reward hacking. \FreshBrew{} is designed to satisfy both of these requirements. The following subsections describe our dataset curation process and the multi-faceted evaluation protocol that defines a successful migration.

\subsection{Dataset Construction}

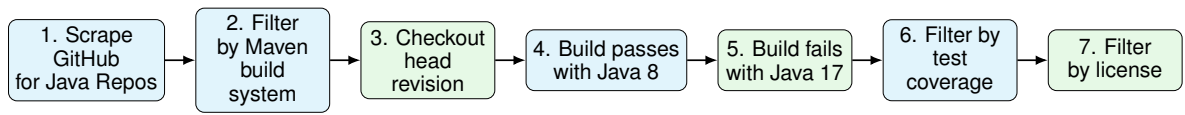
\begin{figure*}[!ht]
  \centering
  \resizebox{0.95\linewidth}{!}{%
    \begin{tikzpicture}[
        node distance = 0.5cm and 0.5cm,
        every node/.style = {font=\sffamily},
        procA/.style = {rectangle, draw, rounded corners,
                        minimum width=2.2cm, minimum height=1cm,
                        align=center, fill=cyan!10},
        procB/.style = {rectangle, draw, rounded corners,
                        minimum width=2.2cm, minimum height=1cm,
                        align=center, fill=limegreen!12},
        arrow/.style = {-{Latex[length=2mm,width=1.6mm]}, thick}
      ]

      % Nodes ----------------------------------------------------
      \node[procA] (scrape)   {1. Scrape\\GitHub\\for Java Repos};
      
      \node[procA, right=of scrape] (maven)  {2. Filter\\by Maven\\build\\system};
      \node[procB, right=of maven]   (checkout) {3. Checkout\\head\\revision};
      \node[procA, right=of checkout] (java8) {4. Build passes\\with Java 8};
      \node[procB, right=of java8]   (java17) {5. Build fails\\with Java 17};
      \node[procA, right=of java17]  (persist) {6. Filter by\\test\\coverage};
      \node[procB, right=of persist] (license) {7. Filter\\by license};

      % Arrows ---------------------------------------------------
      \foreach \i/\j in {scrape/maven,
                         maven/checkout,
                         checkout/java8,
                         java8/java17,
                         java17/persist,
                         persist/license}%
        \draw[arrow] (\i) -- (\j);

    \end{tikzpicture}
  }
  \caption{Automated dataset‐construction pipeline used in this study.}
  \label{fig:dataset-pipeline}
\end{figure*}

To construct our benchmark, we curated a set of Java projects suitable for a migration study through a multi-stage filtering pipeline, as illustrated in Figure~\ref{fig:dataset-pipeline}. Our process is fully automated, ensuring the benchmark can be easily regenerated or extended.

We focused on Maven-based projects as their declarative, XML-based configuration (pom.xml) is more amenable to automated analysis and modification compared to the imperative, code-as-configuration approach of systems like Gradle. 

Our dataset curation process started with 30,000 most popular, by star count, Maven-based Java repositories from GitHub.
From this initial pool, our automated pipeline first confirmed that 6,554 repositories successfully built and passed all tests on Java 8. We then excluded the  projects that also built on Java 17, leaving 1,746 repositories that represent genuine migration tasks. For this set, we enforced quality constraints. Test coverage was calculable for 1,214 of these projects, with 284 meeting our minimum 50\% coverage requirement. Finally, after ensuring each project had a permissive license for accessibility, we arrived at our final dataset of 228 popular repositories, with a median star count of 194 and a minimum of 76.

This construction deliberately targets non-trivial migration tasks rather than a representative sample of all Java projects. In particular, we exclude repositories that already build successfully on JDK 17, thereby removing projects whose migration is trivial or already complete. In addition, because the benchmark is drawn from public, Maven-based GitHub repositories with high test coverage and permissive licenses, the resulting dataset may overrepresent reusable libraries and infrastructure components relative to private application code. The reported success rates should therefore be interpreted as agent performance on a challenging, reproducible migration benchmark, not as an estimate of the overall difficulty of migrating arbitrary Java projects.

% \begin{figure}[!h]
%     \centering
%     \includegraphics[width=0.5\linewidth]{figures/dep_counts.pdf} 
%     \caption{Distribution of the most common dependencies across the repositories in the FreshBrew dataset. The high frequency of foundational libraries like JUnit and Mockito highlights the presence of authentic and non-trivial migration challenges, because testing, mocking, logging, and serialization frameworks frequently constrain dependency and build-configuration choices during migration.}

%     \label{fig:deps_count}
% \end{figure}

% \begin{figure}[!htbp]
%     \centering
%     \includegraphics[width=0.5\linewidth]{figures/license_stats.pdf} \caption{Distribution of open-source licenses across the repositories in the dataset.}
%     \label{fig:license_stats}
% \end{figure}

\begin{figure}[!htbp]
    \centering
    % Subfigure for Dependencies
    \begin{subfigure}[b]{0.48\textwidth}
        \centering
        \includegraphics[width=\textwidth]{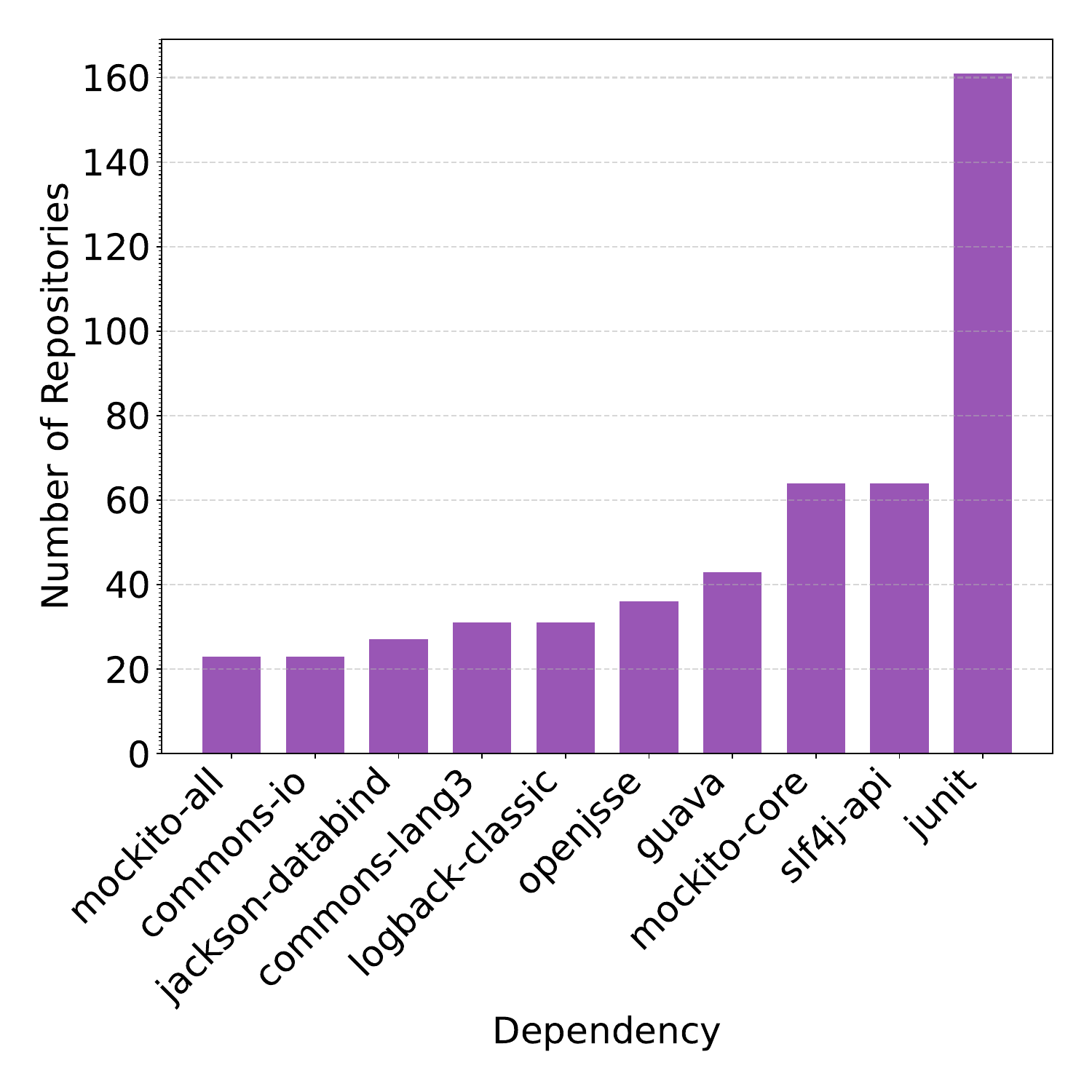}
        \caption{Distribution of common dependencies.}
        \label{fig:deps_count}
    \end{subfigure}
    \hfill % Adds horizontal space between the subfigures
    % Subfigure for Licenses
    \begin{subfigure}[b]{0.48\textwidth}
        \centering
        \includegraphics[width=\textwidth]{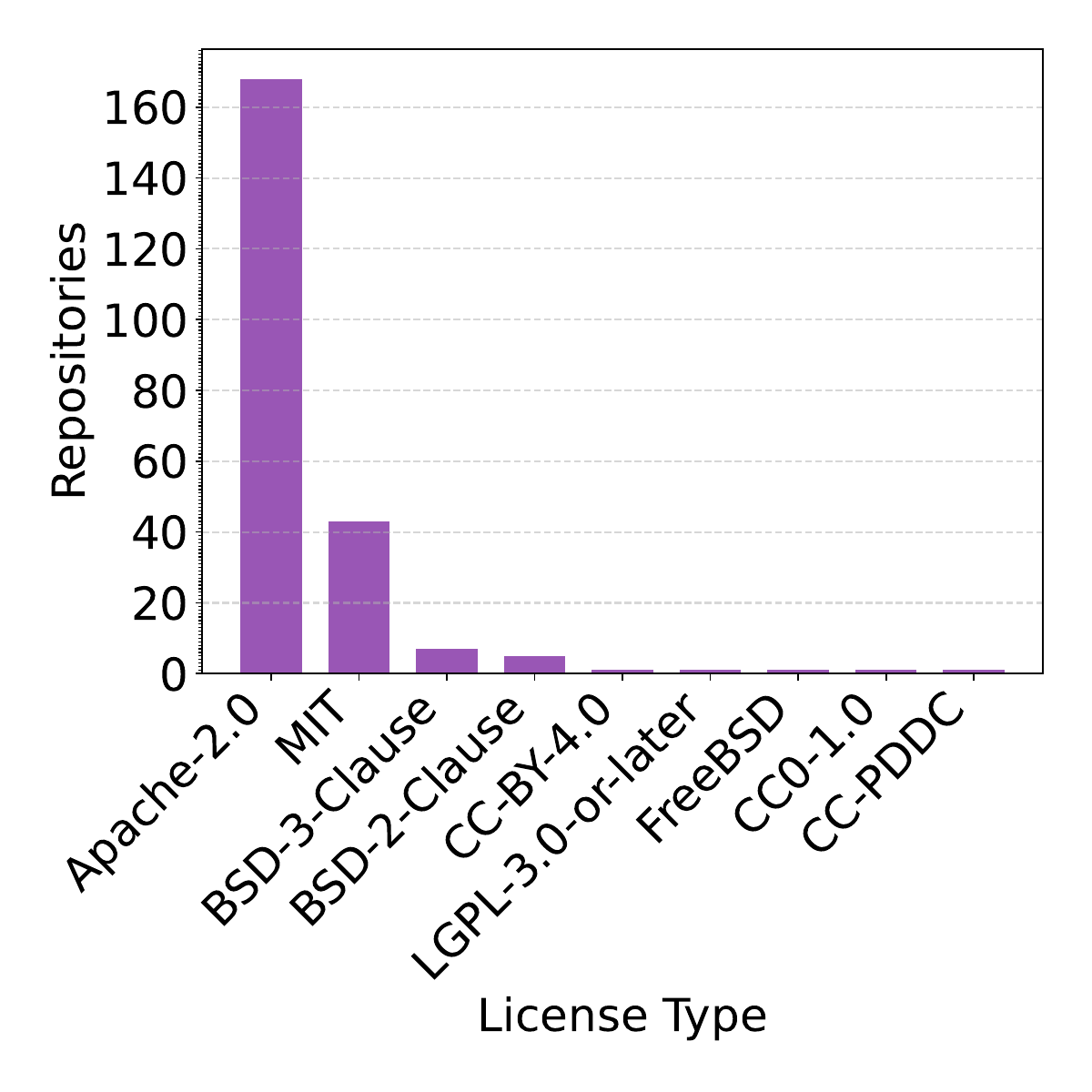}
        \caption{Distribution of open-source licenses.}
        \label{fig:license_stats}
    \end{subfigure}
    
    % Main figure caption
    \caption{Overview of repository statistics in the FreshBrew dataset.}
    \label{fig:dataset_overall_stats}
\end{figure}

Figure~\ref{fig:deps_count} illustrates the distribution of dependencies among the 228 repositories. The results show that the dataset is composed of standard, non-trivial Java projects, with common testing, mocking, logging, and serialization dependencies such as JUnit, Mockito~\cite{Mockito}, SLF4J~\cite{SLF4J}, and Jackson Databind~\cite{JacksonDatabind}. Such dependencies can constrain migration because agents may need to update library versions, Maven plugins, transitive dependencies, or test infrastructure while preserving project behavior.

\begin{figure*}[ht]
    \centering
    \includegraphics[width=0.95\linewidth]{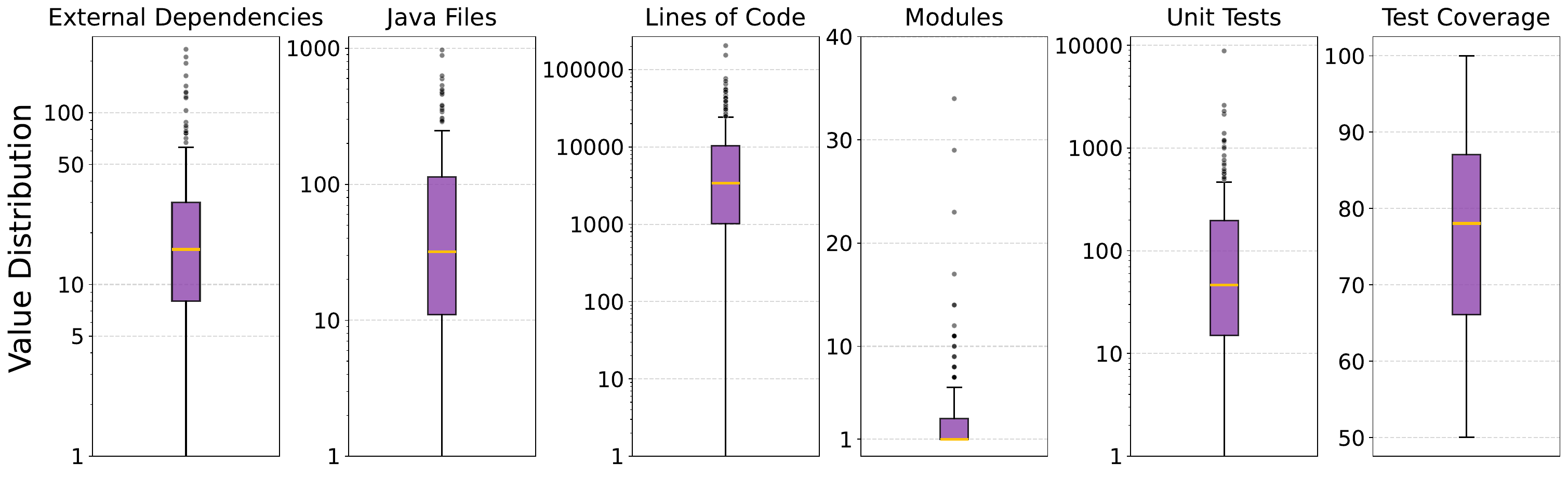} 
    \caption{Distribution of key statistics for repositories in the dataset. The y-axis for metrics with wide ranges (e.g., Lines of Code) is logarithmic to visualize the heavily skewed data. Each plot shows the median (orange line), interquartile range (box), and outliers (dots).}
    \label{fig:dataset_stats}
\end{figure*}

Further statistics of the resulting dataset are presented in Figures~\ref{fig:dataset_stats} and~\ref{fig:license_stats}.

\subsection{Evaluation Protocol}
\label{sec:protocol}
We measure performance on \FreshBrew{} with the metrics outlined below.

\textbf{Overall Success Rate} A migration is considered a success if and only if all of the following conditions are met:
\begin{itemize}
\item \textbf{Compiles}: The migrated project must compile successfully (\texttt{mvn compile}).
\item \textbf{Passes Tests}: All original tests must pass without modification (\texttt{mvn verify}).
\item \textbf{Maintains Coverage.} 
Test coverage is measured using the JaCoCo tool (v0.8.9) with \texttt{LINE} counters, 
aggregated across all Maven modules. A migration is considered successful only if 
the total line coverage does not drop by more than 5\,percentage points relative 
to the original Java 8 baseline. 

\end{itemize}

Enforcing that test line coverage is maintained is a critical safeguard against reward hacking, as it prevents agents from removing either test code or the production code it covers. An agent could achieve a superficially successful migration by simply deleting tests that fail on the new JDK. Similarly, if an agent cannot fix an incompatibility in a specific module of the main application, it might resort to deleting that module to resolve build errors. In either case, there would be a drop in measured line coverage, which would cause the migration to fail our evaluation.

\begin{wrapfigure}{r}{0.5\textwidth}
\vspace{-8mm}
    \begin{center}
      \includegraphics[width=0.48\textwidth]{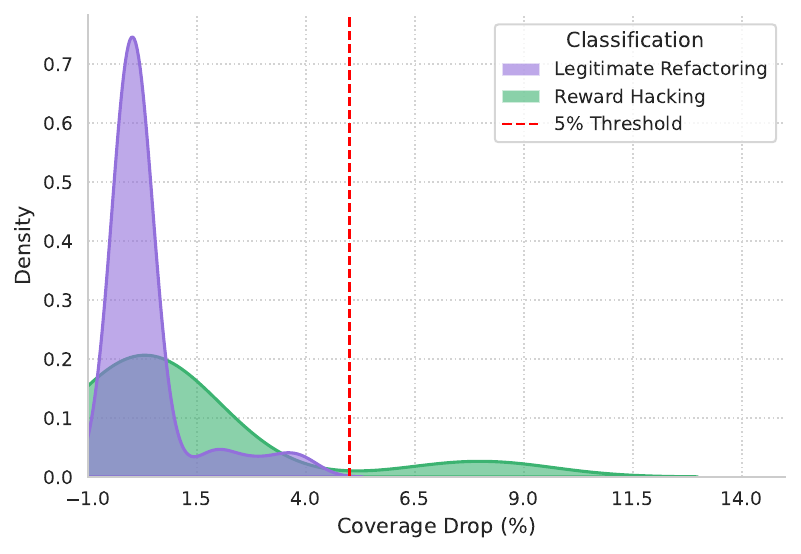}   
    \end{center}
    \caption{Density plot of coverage drops for migrations classified as Legitimate Refactoring versus Reward Hacking. The clear separation between the two distributions supports the 5\% threshold as a conservative boundary for identifying reward hacking.}
    \vspace{-12mm}
    \label{fig:cov_drop}
\end{wrapfigure}

To establish an appropriate threshold, we conducted an empirical audit of 50 migration attempts, classifying each as either "Legitimate Refactoring" or "Reward Hacking". As shown in Figure~\ref{fig:cov_drop}, the analysis reveals a clear distributional separation between the two classes. Legitimate refactorings consistently resulted in coverage drops below 2.5\%, whereas reward hacking attempts showed much larger and more variable drops. %Based on this data, we selected a 5\% threshold as a conservative boundary that effectively distinguishes valid migrations from reward hacking.

Our analysis revealed that coverage drops greater than 5\% were consistently attributable to reward hacking. While many reward hacking instances also occur below this threshold, they are difficult to distinguish from legitimate refactoring using coverage drop alone. We therefore selected 5\% as a conservative threshold to reliably identify a clear subset of reward hacking attempts.

\textbf{Efficiency Metrics}
Beyond correctness, we also measure the efficiency of each agentic migration to understand its practical costs. We focus on the following metrics:

\begin{itemize}
    \item \textbf{Agent Steps.} We record the total number of interaction steps (i.e., thought-action cycles) an agent takes to complete a task. This metric serves as a proxy for the complexity of the agent's solution path. Fewer steps generally indicate a more direct and efficient problem-solving strategy.

    \item \textbf{Cost.} We measure the total cost of using the LLM during a migration run. This metric directly correlates with overall latency. We measure the cost of utilizing each agent by calculating the expense based on the per-token input and output pricing for each LLM, as provided by the \textbf{together.ai\footnote{\url{together.ai/}}} API.

    % \item \textbf{Edit Size.} To quantify the invasiveness of a migration, we calculate the code churn of the agent's final patch. This is measured by the total number of lines of code added and deleted. A smaller, more targeted edit is generally considered higher quality and easier for a human developer to review and maintain. \todo{present this result}
\end{itemize}

\section{Experiments}
\label{sec:experiments}
To demonstrate the capabilities of our benchmark, \FreshBrew{}, we conducted a comprehensive evaluation of seven state-of-the-art large language models and a deterministic migration tool baseline to perform project-level migrations from Java 8 to both Java 17 and Java 21. This section details our experimental setup~(\ref{sec:exp_setup}), reports the migration success rates (\ref{sec:exp_res}), and provides an in-depth analysis of agent behavior and performance~(\ref{sec:exp_anl}). \todo{Also 8->17->21.}
\sloppy

\subsection{Experimental Setup}
\label{sec:exp_setup}
We configured a tool-augmented agent to perform project-level migrations from Java 8 to both Java 17 and Java 21. To provide a point of comparison, we also evaluated OpenRewrite, a  rule-based refactoring tool. This section details our experimental setup, including the agent's environment, models and tools~(\ref{sec:tag}), the OpenRewrite setup~(\ref{sec:or_setup}) and the setup of an experiment to determine the failure modes of the the tool-augmented agent~(\ref{sec:failure_mode}).

\subsubsection{Tool-Augmented agent}
\label{sec:tag}
We conducted experiments with a CodeAct~\cite{wang2024executablecodeactionselicit} agent, as implemented by the smolagents~\cite{smolagents} framework~\footnote{\url{huggingface.co/docs/smolagents/en/reference/agents\#smolagents.CodeAgent}}. To ensure comprehensive coverage, we evaluated a diverse subset of models, including open-weight models, enterprise-grade models, and specialized coding models.

The agent operates in an environment equipped with a set of tools to interact with the file system, build the project, and access external knowledge. The available tools include:
\begin{itemize}
    \item \texttt{read\_file}, \texttt{write\_file}, \texttt{list\_dir}: For basic file system operations.
    \item \texttt{maven\_verify}: A script that executes \texttt{mvn verify} to compile the code and run the full test suite.
    \item \texttt{duckduckgo}: For web search capabilities to find information on libraries or APIs. The tool returns up to 10 search results at a time.
\end{itemize}

The agent was configured to run up to 100 steps and the prompt template is presented in Figure~\ref{fig:agent_prompt_template}. Following~\cite{chen2021evaluatinglargelanguagemodels}, we use a temperature of 0.2 for sampling the models.

\todo{search: duckduckgo (10 results), gemini search and perplexity.}

\todo{complexity analysis of changes needed to migrate.
how many files and loc were edited?}

\subsubsection{Deterministic Baseline with OpenRewrite}
\label{sec:or_setup}
To contextualize the performance of the AI agents, we established a baseline using OpenRewrite, a state-of-the-art deterministic refactoring tool. We evaluated its ability to perform the migration using the composite recipe \url{java.migrate.UpgradeToJava21}~\footnote{\url{docs.openrewrite.org/recipes/java/migrate/upgradetojava21}}. This recipe programmatically applies a series of fine-grained transformations, such as updating Maven compiler settings and replacing deprecated APIs, by operating on a Lossless Semantic Tree (LST) representation of the source code.

For each of the 228 repositories, we attempted to generate an LST and apply the recipe using the Moderne CLI. Due to variations in build configurations and dependency resolution, 69 repositories failed to build an LST. For the remaining 159 repositories, the recipe was applied successfully, and the resulting patches were used for evaluation.

To ensure a direct comparison against the agent-based approaches, these 69 instances where the LST could not be built were considered migration failures. Accordingly, the success rates for OpenRewrite reported in Table~\ref{tab:model_performance} are calculated out of the full dataset of 228 repositories.

We note that OpenRewrite was not intended to be used as an autonomous tool, but rather as a means of saving development time. Therefore, it is reasonable to expect that it would underperform on end-to-end migrations, as compared to AI agents.

\subsubsection{Failure Mode Analysis}
\label{sec:failure_mode}
To qualitatively understand the limitations of the agents, we conducted a failure mode analysis on all unsuccessful migration attempts. 

We employed an LLM-as-Judge approach~\cite{gu2025surveyllmasajudge}, where the Gemini 2.5 Pro~\cite{comanici2025gemini25pushingfrontier} model was prompted to classify the root cause of each failure based on the agent's final 10 steps. We defined a taxonomy of common failure modes, including "Java API Incompatibility," "Dependency Management Failure," "Build Configuration Error," and "Agent Behavioral Failure". An additional "Unknown" category was included for any failures that did not fit the predefined classes; however, all observed failures were classifiable, so this category does not appear in our final analysis. The judge was instructed to select the single best category and provide a brief justification, allowing us to aggregate and quantify the primary reasons for failure for each model.

To ensure the validity of this method, the authors manually reviewed the classifications for 20 randomly sampled failures and found the LLM's reasoning and categorization to be consistent with our assessment in 19 of the 20 cases. This provided us with confidence in the reliability of the overall failure analysis.

\todo{
extended analysis:
distribution of finish step for successful and failed migrations.
same project -> different models, different agent
a couple of examples of maven step status plot.}

\subsection{Experimental Results}
\label{sec:exp_res}
The end-to-end success rates of the OpenRewrite baseline and the seven evaluated models, on the JDK 17 and JDK 21 migration tasks are presented in Table~\ref{tab:model_performance}. 

Overall, we observe a wide variance in performance across the different models, demonstrating that the \FreshBrew{} benchmark poses a significant challenge for modern agentic frameworks. The highest end-to-end success rate on the JDK 17 migration task was achieved by Gemini 2.5 Flash at 52.3\%, while the lowest was DeepSeek-V3 at 10.7\%.

Direct migration from JDK 8 to JDK 21 was generally comparable to, and for most models slightly harder than, direct migration from JDK 8 to JDK 17. These results should be interpreted carefully: both tasks start from the same Java 8 repository revision, so the JDK 21 task measures a direct leap from Java 8 rather than a staged migration through JDK 17. In practice, many projects migrate incrementally, and a repository that has already been adapted to JDK 17 may require fewer additional changes to move to JDK 21. Evaluating such staged migrations is an important direction for future work. This comparison is discussed in more detail in Section~\ref{sec:target_ver_comp}.

% \begin{table*}[!htbp]
% \centering
% \begin{tabular}{
%     l    
%     c c     
%     > {\columncolor{HighlightGray}}c    
%     c c 
%     > {\columncolor{HighlightGray}}c}
% \toprule
% \textbf{Model} & \multicolumn{3}{c}{\textbf{JDK 17 Success Rate}} & \multicolumn{3}{c}{\textbf{JDK 21 Success Rate}} \\
% \cmidrule(lr){2-4} \cmidrule(lr){5-7}
% & \textbf{Compilation} & \textbf{Tests} & \textbf{Cov Guard} & \textbf{Compilation} & \textbf{Tests} & \textbf{Cov Guard}\\
% \midrule
% Gemini 2.0 Flash & 71.8\% & 53.7\% & 44.8\% & 68.2\% & 47.5\% & 35.9\% \\
% Gemini 2.5 Flash & 80.8\% & 66.5\% & 56.5\% & 77.3\% & 60.0\% & 52.4\% \\
% Gpt 4.1 & 82.0\% & 55.7\% & 47.0\% & 66.7\% & 43.4\% & 39.1\% \\
% Gpt 4o & 63.2\% & 32.0\% & 24.8\% & 52.6\% & 24.1\% & 19.7\% \\
% O3 Mini & 52.2\% & 23.9\% & 27.8\% & 40.4\% & 8.3\% & 4.5\% \\
% \bottomrule
% \end{tabular}
% \caption{Performance on Migration}
% \label{tab:model_performance}
% \end{table*}

\begin{table*}[htbp]
\centering
\resizebox{\textwidth}{!}{
\begin{tabular}{
    l    
    c c     
    > {\columncolor{HighlightGray}}c    
    c c 
    > {\columncolor{HighlightGray}}c}
\toprule
\textbf{Model / Method} & \multicolumn{3}{c}{\textbf{JDK 17}} & \multicolumn{3}{c}{\textbf{JDK 21}} \\
\cmidrule(lr){2-4} \cmidrule(lr){5-7}
& \textbf{Compilation} & \textbf{Tests} & \textbf{\shortstack[c]{Overall\\Success\\Rate}} & \textbf{Compilation} & \textbf{Tests} & \textbf{\shortstack[c]{Overall\\Success\\Rate}}\\
\midrule
\multicolumn{7}{@{}l}{\textbf{Rule-Based Systems}} \\
OpenRewrite & 54.4\% & 7.0\% & 7.0\% & 57.5\% & 7.5\% & 7.5\% \\
\hspace*{1em}\textit{on projects w/ successful LST build (159/228)} & \textit{78.0\%} & \textit{10.1\%} & \textit{10.1\%} & \textit{82.4\%} & \textit{10.7\%} & \textit{10.7\%} \\
\addlinespace[0.5em]
\midrule
\multicolumn{7}{@{}l}{\textbf{Open-Weights Models}} \\
DeepSeek-V3~\cite{deepseekai2025deepseekv3technicalreport}& 55.9\% & 13.7\% & 10.7\% & 50.4\% & 21.7\% & 12.4\% \\
Qwen3 ~\cite{yang2025qwen3technicalreport} & 59.2\% & 18.0\% & 15.9\% & 43.0\% & 14.5\% & 12.8\% \\
\addlinespace[0.5em]
\multicolumn{7}{@{}l}{\textbf{Enterprise Models}} \\
Gemini 2.5 Flash & 79.8\% & 63.2\% & 52.3\% & 75.4\% & 58.3\% & 49.8\% \\
%Gemini 2.5 Flash~\cite{comanici2025gemini25pushingfrontier} & 80.0\% & 64.1\% & \textbf{54.3}\% & 77.9\% & 60.8\% & \textbf{52.4\%} \\
GPT-4.1 ~\cite{gpt4_1_openai_2025} & 76.8\% & 55.7\% & 47.1\% & 70.6\% & 49.1\% & 44.2\% \\
GPT-4o ~\cite{gpt4o_system_card_2024} & 64.0\% & 34.2\% & 30.9\% & 57.0\% & 28.1\% & 24.9\% \\
o3-mini ~\cite{openai_o3_system_card} & 52.2\% & 36.9\% & 27.8\% & 40.4\% & 8.3\% & 4.5\% \\

\addlinespace[0.5em]
\multicolumn{7}{@{}l}{\textbf{Specialized Coding Models}} \\
Arcee AI Coder-Large~\cite{arceeModelSelection} & 51.3\% & 22.8\% & 21.1\% & 57.5\% & 21.7\% & 20.2\% \\
%Adk Gemini 2.5 Flash & 76.8\% & 65.5\% & 60.7\% & 70.1\% & 47.4\% & 42.7\% \\
%Adk Gemini 2.5 Pro & 78.3\% & 58.4\% & 49.1\% & 72.8\% & 56.3\% & 48.1\% \\

\addlinespace[0.5em]
\multicolumn{7}{@{}l}{\textbf{Enterprise Models / ADK}} \\
Gemini 2.5 Flash & 71.1\% & 52.6\% & 48.4\% & 66.2\% & 41.7\% & 37.2\% \\
Gemini 2.5 Pro & 77.6\% & 56.6\% & 47.5\% & 71.8\% & 53.7\% & 46.6\% \\

\bottomrule
\end{tabular}}
\caption{Performance of AI models on the JDK~17 and JDK~21 migration tasks.
Success is measured in three stages: \textit{Compilation} (the project builds on the target JDK),
\textit{Tests} (all original tests pass unmodified), and \textit{Overall Success Rate},
which additionally requires that test line coverage does not drop by more than 5\,pp relative
to the Java~8 baseline (Section~\ref{sec:protocol}). This safeguard prevents agents from reward-hacking
by deleting code or tests.}

%\caption{\textbf{Performance of AI models on the JDK 17 and JDK 21 migration tasks.} Success is measured at three stages: successful \texttt{Compilation}, passing all \texttt{Tests}, and the final \texttt{Overall Success Rate}. A migration is only considered an "overall success" if it satisfies all evaluation criteria, including compiling, passing the full test suite, and maintaining test coverage. For OpenRewrite, we include also the success rate on projects that could be parsed with AST.}
\label{tab:model_performance}
\end{table*}

\subsection{Experiment Analysis}
\label{sec:exp_anl}
While the overall success rates provide a high-level view of model performance, a deeper analysis is required to understand the underlying behaviors and challenges. In this section, we dissect the experimental outcomes to uncover key insights. We analyze agent traces to compare their problem-solving efficiency, investigate how project complexity impacts performance, categorize the root causes of unsuccessful migrations, and present illustrative case studies to highlight the practical challenge of reward hacking. 

\subsubsection{Agent Trace Analysis}
\begin{figure}[!htbp]
    \centering
    % Subfigure for Step Counts
    \begin{subfigure}[b]{0.48\textwidth}
        \centering
        \includegraphics[width=\textwidth]{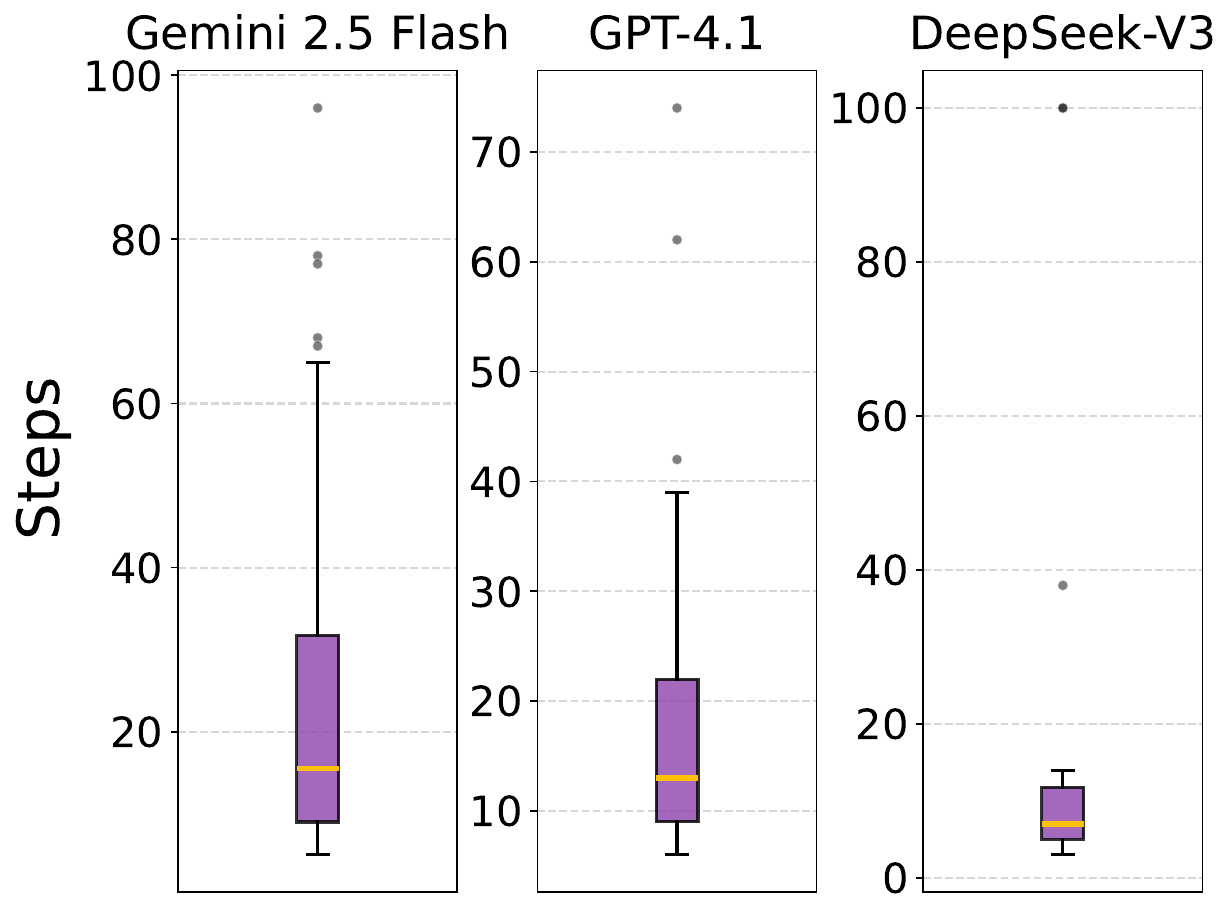}
        \caption{Comparison of step counts.}
        \label{fig:trace_steps_hist}
    \end{subfigure}
    \hfill % Adds horizontal space
    % Subfigure for Costs
    \begin{subfigure}[b]{0.48\textwidth}
        \centering
        \includegraphics[width=\textwidth]{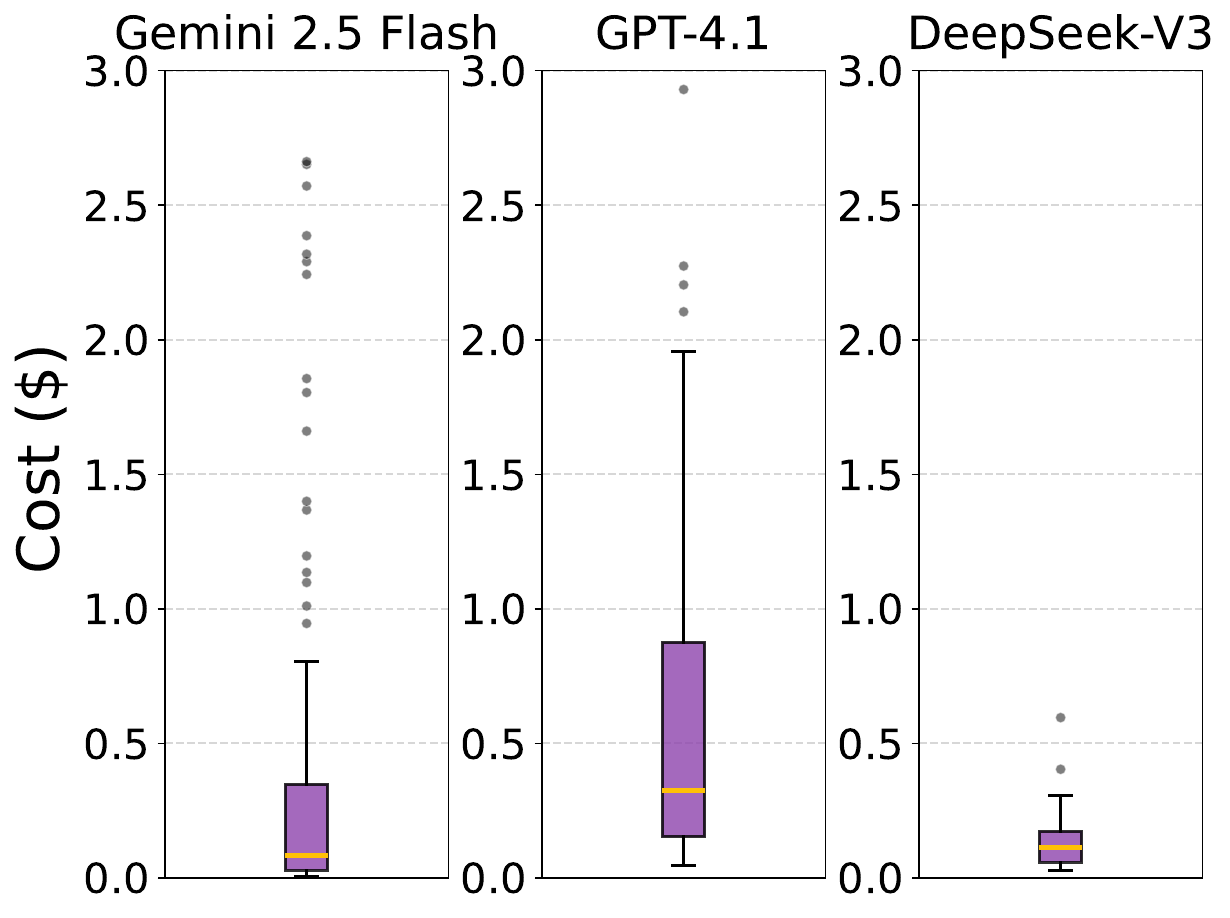}
        \caption{Comparison of migration costs.}
        \label{fig:trace_tokens_hist}
    \end{subfigure}
    
    % Main figure caption
    \caption{A comparison of step count and cost distributions for successful migrations to Java 17. The boxplots show the distribution for (a) the number of steps and (b) the associated costs.}
    \label{fig:migration_stats}
\end{figure}

The distributions of step counts and cost of successful migrations are presented in Figures~\ref{fig:trace_steps_hist} and ~\ref{fig:trace_tokens_hist}. For a clearer analysis of agent efficiency, these figures focus on a representative subset of the models: \textbf{Gemini 2.5 Flash} and \textbf{GPT-4.1} as leading proprietary models, and \textbf{DeepSeek-V3} as a top-performing open-weight model.

% \begin{figure}[!ht]
%     \centering
%     \includegraphics[width=0.5\linewidth]{figures/step_boxplots.pdf} 
%     \caption{A comparison of step count distributions for successful migrations to Java 17.}
%     \label{fig:trace_steps_hist}

% \end{figure}

% \begin{figure}[!htbp]
%     \centering
%     \includegraphics[width=0.5\linewidth]{figures/costs_boxplots.pdf} 
%     \caption{A comparison of cost distributions for successful migrations to Java 17.}
   
%     \label{fig:trace_tokens_hist}
% \end{figure}

% \begin{figure}[!ht]
%     \centering
%     \includegraphics[width=0.5\linewidth]{figures/step_success_correlation_grouped.pdf} 
%     \caption{\textbf{A comparison of model success rates on Java 17 migration, binned by task complexity}. The x-axis represents the number of agent steps required to solve a problem. The y-axis shows the success rate for each model within that bin.}
   
%     \label{fig:step_success_correlation}
% \end{figure}

% \begin{figure}[!ht]
%     \centering
%     \includegraphics[width=0.5\linewidth]{figures/failure_modes_comparison.pdf} 
%     \caption{\textbf{Distribution of failure modes on Java 17 Migration for each evaluated model, as determined by an LLM-as-Judge analysis}. Each failure category is sorted in descending order based on its highest prevalence across the models.}
%     \label{fig:failure_modes}
% \end{figure}

Our analysis of agent steps in Figure~\ref{fig:trace_steps_hist} shows that the models employ distinct approaches. \textbf{DeepSeek-V3} appears to follow a highly direct strategy, resolving successful migrations with the lowest median number of steps (around 5). \textbf{GPT-4.1} represents a balanced approach with a median of approximately 13 steps. In contrast, \textbf{Gemini 2.5 Flash} engages in a more extensive exploratory process, requiring a higher median of ~17 steps and showing the widest variability.

In regard to cost efficiency, (Figure~\ref{fig:trace_tokens_hist}) shows the cost profiles for successful migrations. \textbf{DeepSeek-V3} is the most economical, with a low median cost and tight distribution. Conversely, \textbf{GPT-4.1} has the most variable typical costs. \textbf{Gemini 2.5 Flash} also has a low median cost but is distinguished by a long tail of high-cost outliers.
%Gemini 2.5 Flash exhibited a high variability compared to the other models. DeepSeek-V3 was highly efficient.

Figure~\ref{fig:step_success_correlation} provides a granular analysis of model performance by segmenting the success rate according to the number of agent steps required for each task. This metric serves as a proxy for procedural complexity, offering insights into how each model's effectiveness changes as problems become more difficult.

A notable finding across all models is that peak performance is achieved not on the simplest tasks (1-5 steps), but on those of moderate complexity requiring 6-10 steps. This suggests a potential "sweet spot" where problems are sufficiently involved to engage the models' reasoning capabilities without becoming intractable.

Among the models evaluated, \textbf{Gemini 2.5 Flash} demonstrates the most robust performance profile. After achieving a near-perfect success rate in the 11-20 step bin, its performance degrades more gradually than its competitors, establishing it as the most effective model for highly complex tasks requiring over 20 steps. 

In summary, our trace analysis reveals that the choice of a backend model for agentic migration involves significant trade-offs in cost and speed versus success rate.

\subsubsection{\texorpdfstring{Direct Java 8$\rightarrow$17 vs.\ Direct Java 8$\rightarrow$21 Migration}{Direct Java 8 to 17 vs. Direct Java 8 to 21 Migration}}
\label{sec:target_ver_comp}

\begin{wrapfigure}{r}{0.5\textwidth}
    \vspace{-5mm}
    \begin{center}
    \includegraphics[width=0.48\textwidth]{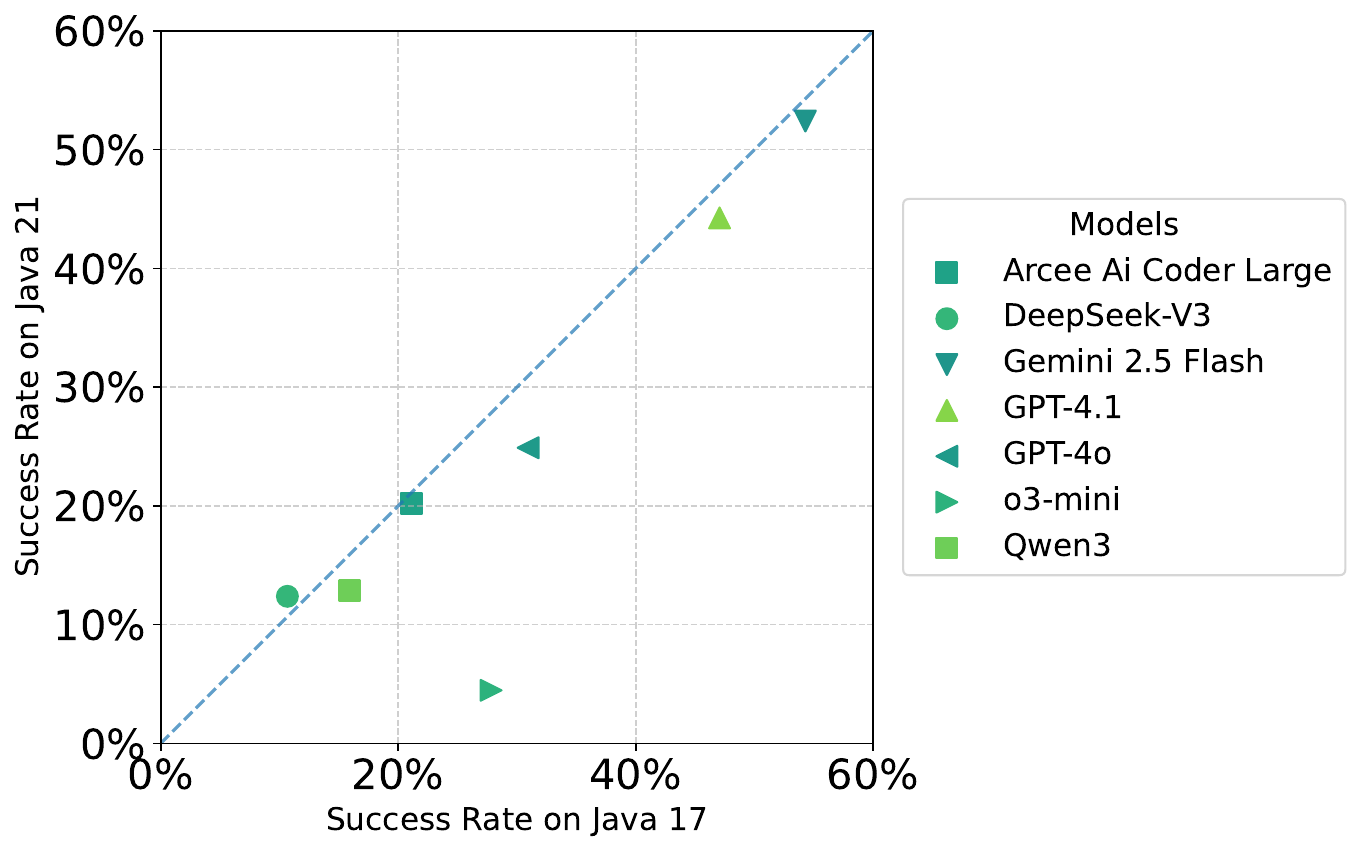} 
    \end{center}
    \caption{\textbf{A scatter plot comparing the success rates of various models on JDK 17 versus JDK 21 migration tasks}. The dashed line indicates equal performance on both tasks.}
    \vspace{-10mm}
    \label{fig:target_ver_scatter}
\end{wrapfigure}

To directly compare model performance across the two migration tasks, we visualized the overall success rates for the JDK 17 and JDK 21 targets in a scatter plot (Figure~\ref{fig:target_ver_scatter}). This visualization allows for an immediate assessment of model consistency and the relative difficulty of the tasks.

The analysis of Figure~\ref{fig:target_ver_scatter} shows a strong positive correlation in model performance between the JDK 17 and JDK 21 migration tasks. Most models perform either equally well or slightly worse on JDK 21, clustering near or below the line of parity. DeepSeek-V3 is a mild outlier, achieving a small improvement from 10.7\% to 12.4\% on JDK 21. Given the study's single-run ($n=1$) design, these small differences may reflect model stochasticity, suggesting that for most models the two tasks present a largely comparable level of difficulty. A notable exception was o3-mini, whose success rate fell sharply from 27.8\% to 4.5\%, indicating that some models are significantly less resilient to the specific changes in the newer Java version.

Figure~\ref{fig:target_ver_scatter} compares two direct migrations from the same Java 8 starting point. It does not measure a staged workflow in which a project is first successfully migrated to JDK 17 and then migrated from JDK 17 to JDK 21. This distinction is important because many Java 9--17 compatibility issues (such as dependency upgrades, build-plugin updates, class-loader assumptions, and access to JDK internals) would already have been resolved by the time a project begins a 17$\rightarrow$21 migration. A natural extension of \FreshBrew{} is therefore a third task that evaluates staged JDK 17$\rightarrow$21 migrations starting from successful JDK 17 outputs.
%The results reveal two key trends. First, there is a clear positive correlation; models that performed well on the JDK 17 migration generally performed well on the JDK 21 migration. Second, nearly all models fall below the diagonal line of equal performance. This positioning indicates that the migration to JDK 21 was more challenging for the entire suite of evaluated models.

\subsubsection{Model Performance vs. Project Complexity}

\begin{figure*}[!htbp]
    \centering
    \includegraphics[width=0.95\linewidth]{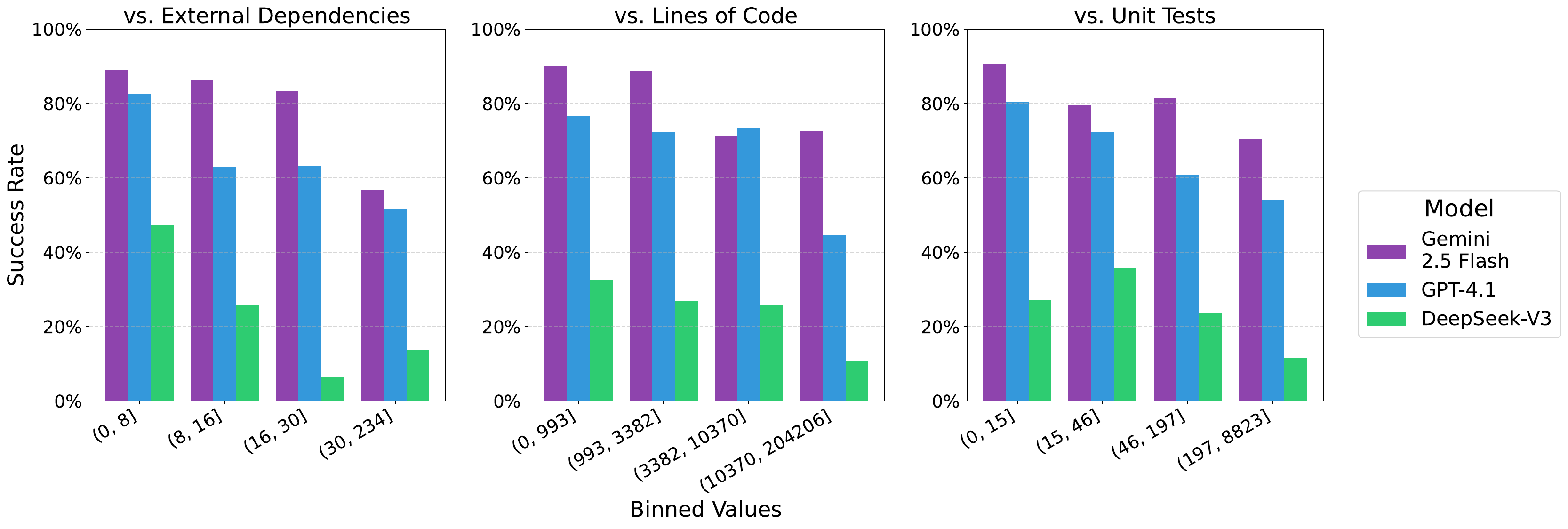} 
    \caption{\textbf{Model Performance on the Java 17 Migration Task as a Function of Project Complexity}. The migration success rate for each model is plotted against quartiles of different complexity metrics. For all models, performance consistently degrades as the number of external dependencies, lines of code, and unit tests increases, demonstrating that the benchmark effectively measures challenges posed by real-world project complexity.}
    \label{fig:perf_vs_comp}
\end{figure*}

We analyzed model performance across bins of varying project complexity. Figure~\ref{fig:perf_vs_comp} shows a clear trend: for all models, the migration success rate consistently decreases as project complexity (measured by dependencies, lines of code, and number of tests) increases. 

\subsubsection{Case Studies in Reward Hacking}

The following case studies illustrate instances of reward hacking wherein an AI agent had optimized for the proximate reward signal of a successful build, even when its actions undermined the primary goal of a functional migration.

\paragraph*{Case Study 1: Targeted Test Exclusion}
\todo{More case studies from other models}
\

\noindent\textbf{Repository:} \texttt{DaisyDiff/DaisyDiff}

\noindent\textbf{Model:} \texttt{Gemini 2.5 Flash}

\noindent\textbf{Target Java Version:} \texttt{17}

\noindent\textbf{Situation:} The failures were isolated to two specific unit tests which exhibited incompatibilities with the new environment.

\noindent\textbf{Agent's Action:} Rather than correcting the code-level incompatibilities, the agent modified the \texttt{pom.xml} file to explicitly \textbf{exclude the two failing tests} from the build lifecycle via the maven-surefire-plugin.

\paragraph*{Case Study 2: Ignoring Internal Runtime Failures}
\

\noindent\textbf{Repository:} \texttt{BottegaIT/ddd-leaven-v2}

\noindent\textbf{Model:} \texttt{o3-mini}

\noindent\textbf{Target Java Version:} \texttt{17}

\noindent\textbf{Situation:} The migration to new Spring and Java versions caused a critical runtime error, \url{org.springframework.beans.factory.NoSuchBeanDefinitionException}, for a core business component (\texttt{OrderShipmentStatusTrackerSaga}). This exception was triggered during a test run, but the application's internal event publisher caught the error and logged it, which allowed the JUnit test itself to complete successfully.

\noindent\textbf{Agent's Action:} The agent correctly identified and resolved all compilation errors by updating dependencies in the \texttt{pom.xml} file. However, it accepted the successful test result from the build tool as a final success, \textbf{ignoring the critical runtime error logged during the test's execution}. This resulted in a superficially correct migration where a key piece of business logic was non-functional, a failure only made visible by the corresponding drop in test coverage.

\paragraph*{Case Study 3: Bypassing Tests Broken by JVM Changes}
\

\noindent\textbf{Repository:} \texttt{scijava/native-lib-loader}

\noindent\textbf{Model:} \texttt{GPT-4.1}

\noindent\textbf{Target Java Version:} \texttt{17}

\noindent\textbf{Situation:} A key test, \texttt{NativeLoaderTest}, assumed that the system/application class loader was a \texttt{URLClassLoader} and used reflection to invoke the protected \texttt{URLClassLoader.addURL} method in order to add a JAR at run time. This assumption held for the tested Java 8 environment, but it is not part of the Java SE specification. Since JDK 9, the application class loader is no longer a \texttt{URLClassLoader} but an internal class-loader implementation~\citep{OracleJDKMigrationGuide}. Thus, the failure is caused by reliance on a class-loader implementation detail and reflective access to a non-public method, not by removal of \texttt{URLClassLoader.addURL}.

\noindent\textbf{Agent's Action:} Instead of adapting the test to use a modern approach, the agent wrapped the failing reflective call in a conditional block. It then added logic to the test itself that causes it to \textbf{silently skip its own execution} on modern Java versions. While this allowed the build to pass, it effectively disabled the test, leaving the corresponding production code uncovered.

\subsubsection{Analysis of Failure Modes}
\begin{figure}[htbp]
    \centering
    % Subfigure for Success Correlation
    \begin{subfigure}[b]{0.48\textwidth}
        \centering
        \includegraphics[width=\textwidth]{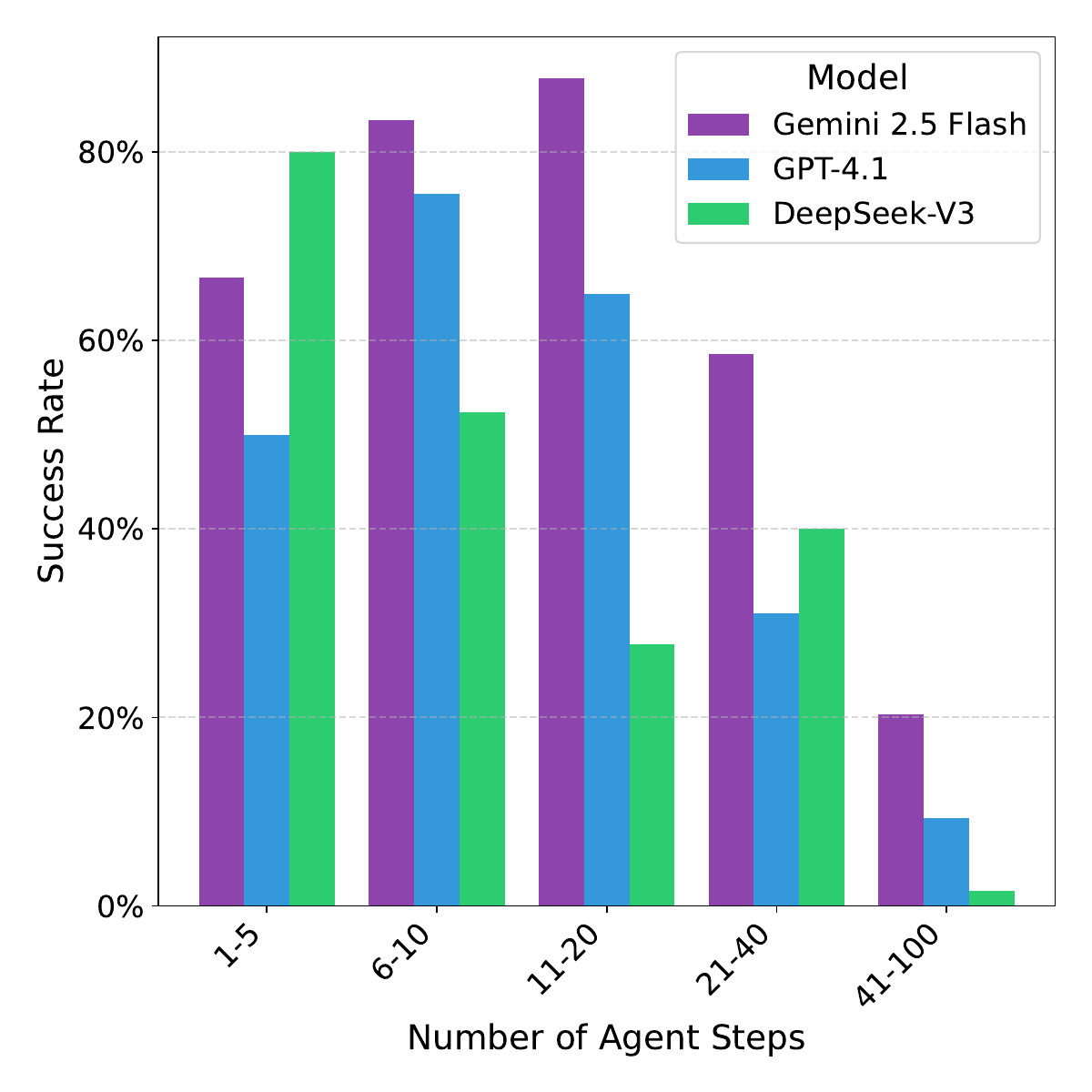}
        \caption{Success rates by task complexity.}
        \label{fig:step_success_correlation}
    \end{subfigure}
    \hfill % Adds horizontal space
    % Subfigure for Failure Modes
    \begin{subfigure}[b]{0.48\textwidth}
        \centering
        \includegraphics[width=\textwidth]{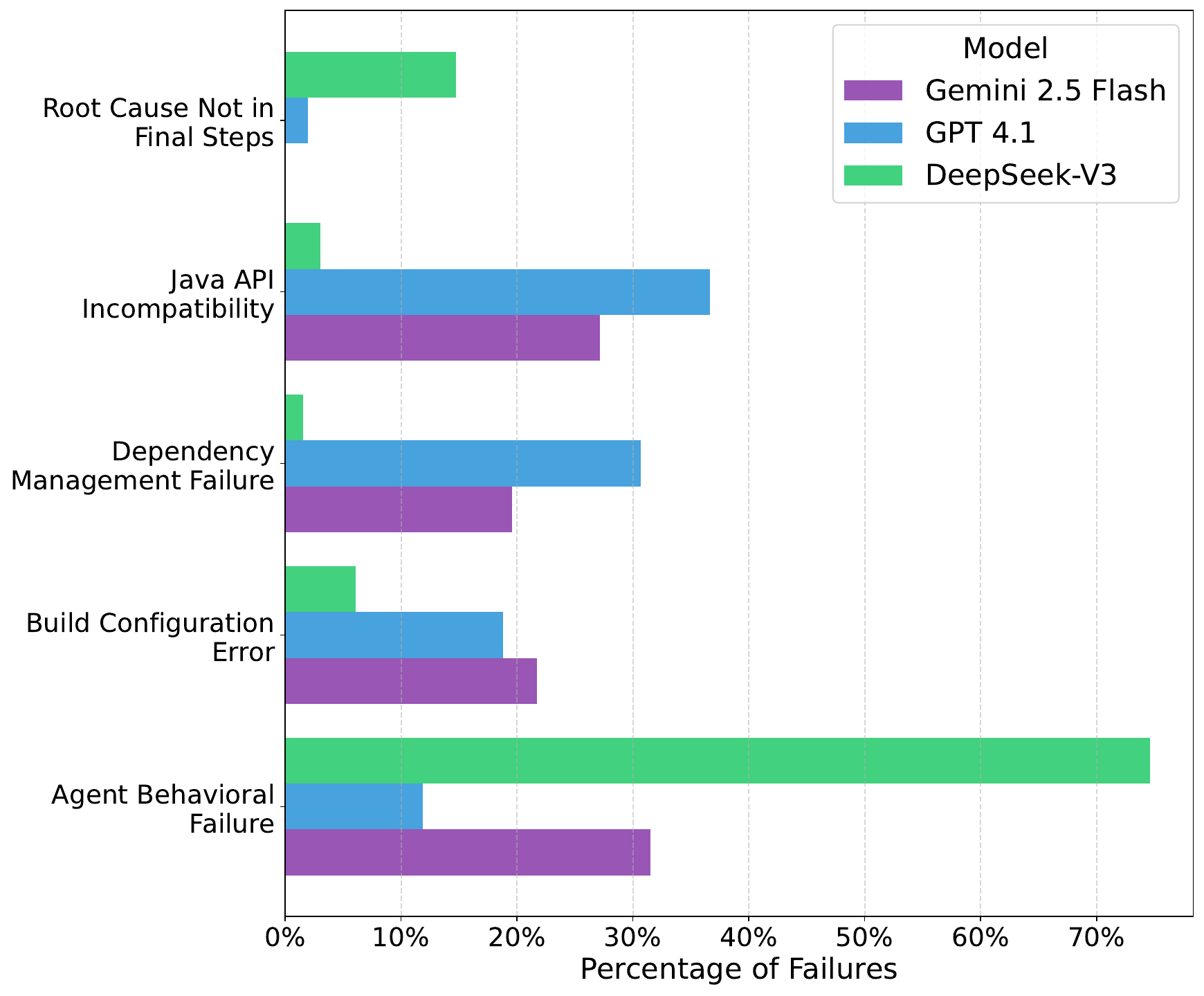}
        \caption{Distribution of failure modes.}
        \label{fig:failure_modes}
    \end{subfigure}
    
    % Main figure caption
    \caption{\textbf{Analysis of model performance on Java 17 migration.} (a) A comparison of model success rates binned by task complexity, measured in agent steps. (b) The distribution of failure modes for each model, determined by an LLM-as-Judge analysis.}
    \label{fig:performance_analysis}
\end{figure}
To understand the limitations of current agents beyond binary success rates, we performed a qualitative analysis on all unsuccessful runs. Using an LLM-as-judge, we categorized each failure based on the agent's final steps. Figure~\ref{fig:failure_modes} presents a comparative breakdown of these failure modes, highlighting the distinct behavioral profiles of each model.

The analysis reveals that \textbf{Agent Behavioral Failure} - where agents get stuck in repetitive loops, hallucinate commands, or fail to make productive edits - is a common issue overall. It is particularly pronounced for \textbf{DeepSeek-V3}, which saw over 70\% of its failures fall into this category.

In contrast, \textbf{Gemini 2.5 Flash} and \textbf{GPT-4.1}, while still susceptible to behavioral issues, failed more frequently due to deeper technical challenges. Both models show a significant percentage of failures in \textbf{Java API Incompatibility} and \textbf{Dependency Management Failure}. This suggests that as models become more capable at basic agentic tasks (like editing files and running commands), their primary bottleneck shifts to the complex reasoning required to resolve breaking API changes and intricate dependency conflicts. For example, \textbf{GPT-4.1} struggled mostly with Java API incompatibility issues, while \textbf{Gemini 2.5 Flash}'s failures were more evenly spread across behavioral, API, and dependency challenges. %This granular analysis provides a clearer picture of the current frontier in automated code migration, pointing to specific areas for future improvement.

\subsubsection{Limitations of Deterministic Baselines}
Our baseline experiment with \textbf{OpenRewrite} highlights a fundamental limitation of rule-based systems in complex migration tasks. \textbf{OpenRewrite} operates deterministically; it can only apply transformations for which an explicit rule exists. It is not designed to handle unforeseen challenges, such as a critical third-party library that is incompatible with the target Java version and has no clear, drop-in replacement.

In such cases, the tool correctly completes its prescribed refactoring but leaves the remaining, more complex problem for a human developer to solve. This contrasts sharply with the goal of agentic systems, which are designed to tackle these ambiguous, open-ended problems by searching for solutions and attempting novel code modifications. This distinction is critical: while rule-based tools excel at predictable refactoring, they cannot fully automate migrations that require creative problem-solving or dependency-level changes outside their predefined rules.

\section{Limitations}

Our benchmark's external validity is subject to two primary limitations:

\begin{itemize}
    \item \textbf{Focus on Maven:} \FreshBrew{}  currently includes only Maven-based projects. This was a pragmatic choice, as Maven's standardized, declarative format enabled the creation of a robust, automated curation and evaluation pipeline. Extending this to other systems like Gradle is a challenge due to the complexity and variability of their code-based build scripts. Unlike Maven’s declarative XML, Gradle’s imperative build scripts are executable code, which present a more complex program modification challenge. Supporting such build systems remains an important goal for future work.
    \item \textbf{Representativeness of Open-Source Data and Library Skew:} Our dataset's use of public GitHub repositories is a limitation. Public Maven repositories that are popular, permissively licensed, and well-tested may skew toward reusable libraries, frameworks, plugins, and infrastructure components rather than deployed application code. This matters because libraries often interact more directly with build tooling, class loading, reflection, serialization, mocking frameworks, and other lower-level APIs, making them harder to migrate than many application codebases. Moreover, the pipeline intentionally excludes projects that already build on JDK 17, so \FreshBrew{} should be viewed as a benchmark of non-trivial migration tasks rather than a representative sample of all Java migrations. Enterprise systems introduce a different set of challenges, including stale private dependencies, forked libraries, monorepo build systems, and governance gates, which may not be fully captured by public GitHub projects.
    \item \textbf{Selection Bias} While our focus on high-coverage, permissively licensed projects introduces a selection bias, these choices were necessary trade-offs. The high test coverage is a core requirement for our reward-hacking detection protocol, and permissive licenses are an ethical prerequisite for building a public benchmark.  Consequently, our findings on performance are most applicable to the domain of well-maintained, robustly tested software projects.
\end{itemize}

\textbf{Threats to Experimental and Construct Validity:} 
\begin{itemize} 
\item \textbf{Single Generation Pass:} Our study reports results from a single generation pass ($n=1$) per scenario, using a low sampling temperature to favor deterministic outputs. This is a standard practice in large-scale evaluations to ensure reproducibility and manage computational cost, but it does expose a threat~\cite{zhang2024autocoderoverautonomousprogramimprovement, zheng2025makeslargelanguagemodels, jain2025testgenevalrealworldunit}. We chose this approach due to the significant computational cost of running large-scale agentic evaluations on the full dataset. While multiple runs would provide a measure of variance, we hypothesize that our results are representative, as we used a low sampling temperature and a simple, user-aligned prompt to favor deterministic behavior.
\item \textbf{Fixed Prompt:} Our study has a specific limitation separate from the benchmark itself: the use of a single, fixed prompt template (presented in Figure~\ref{fig:agent_prompt_template}) for all agents. The performance rates we report are consequently tied to this specific set of instructions. We did not perform prompt engineering, and it is possible that agent performance could change with more optimized prompts. This is a limitation of our study's methodology, not of the \FreshBrew{} benchmark, which can be used with any agent or prompt configuration.
\item \textbf{The Test Coverage Heuristic:} Our evaluation protocol defines a successful migration as one where test line coverage does not drop by more than 5 percentage points. This threshold was chosen as a balanced heuristic to distinguish legitimate refactoring from reward hacking. A stricter rule could unfairly penalize valid code changes, while a more lenient one could fail to prevent reward hacking. While we validated this choice on a random sample of migration attempts, this heuristic may not be universally optimal for every project or migration context.
\item \textbf{Migration Path Scope:} \FreshBrew{} currently evaluates direct migrations from JDK 8 to JDK 17 and from JDK 8 to JDK 21. This setup is useful because both tasks share a common Java 8 starting point, but it does not model staged migration workflows. In practice, developers often migrate incrementally, for example by first moving from JDK 8 to JDK 17 and then from JDK 17 to JDK 21. A staged JDK 17$\rightarrow$21 benchmark would help isolate the difficulty of moving between modern JDK releases after the major Java 9--17 compatibility issues have already been addressed.
\end{itemize}
\label{sec:limitations}
\section{Conclusion}

In this paper, we address a key challenge at the intersection of AI and software engineering: the reliable evaluation of autonomous agents on complex, repository-level code migration tasks. While prior and parallel benchmarks have focused on the migration problem itself, they were not designed to handle the unique failure modes of AI agents, such as reward hacking.

To fill this gap, we introduce \FreshBrew{}, the first benchmark specifically designed for evaluating agentic Java migrations. Our work presents a threefold contribution:

\begin{itemize}
    \item \textbf{A curated, high-coverage dataset}: We provide a collection of real-world Java projects that are guaranteed to build on JDK 8 but fail on modern JDKs, with each project having significant test coverage to allow for meaningful evaluation.
    
    \item \textbf{A robust evaluation protocol}: We introduce a multi-faceted evaluation method where success is determined not just by compilation and passing tests, but also by maintaining test coverage. This protocol is specifically designed to protect against reward hacking, ensuring a more precise measure of an agent's migration capabilities.
    
    \item \textbf{An empirical study of AI agents}: We present a comprehensive evaluation of state-of-the-art, LLM-based agents, offering insights into their performance, behaviors, and limitations when performing Java migration tasks.
\end{itemize}

Our experiments using \FreshBrew{} yield  insights into the current state of AI agents. We find that, on this challenging subset of non-trivial Java 8$\rightarrow$17 migration tasks, leading models such as Gemini 2.5 Flash can achieve a promising success rate of 52.3\%, while performance and cost remain highly variable across models. Our protocol has uncovered that a significant portion of apparent successes would have been classified as reward hacking without integrity checks, underscoring the critical importance of evaluating agents with specialized tools. 

Future extensions should broaden the benchmark beyond public Maven repositories, characterize the library/application composition of the dataset, and add staged migration tasks such as JDK 17$\rightarrow$21.

By releasing \FreshBrew{} to the community, we aim to provide a robust and extensible platform to drive progress in AI-driven modernization, ensuring the next generation of software engineering agents are not only effective but also reliable and trustworthy.

\label{sec:conclusion}

\section*{Data and Code Availability}

All data and code used in this study are publicly available. 
The \FreshBrew{} benchmark dataset, including the curated repository list, 
evaluation scripts, and agent prompts, is released under a permissive open-source 
license at \url{github.com/mrcabbage972/freshbrew}. 
Each experiment in this paper can be reproduced using the configuration files 
and evaluation pipeline provided in the repository. 
All third-party Java projects included in the dataset are sourced from 
public GitHub repositories under compatible open-source licenses as detailed in 
Section~\ref{sec:benchmark}.

\section{Ethics Statement and Broader Impact}

This study uses only publicly available, permissively licensed open-source
repositories from GitHub. All projects in the \FreshBrew{} dataset were selected
with explicit license checks to exclude non-permissive or proprietary material.
The benchmark’s evaluation protocol is designed to discourage reward hacking and
other behaviors that could degrade software quality or safety. No personal or
user-generated data are included, and no human subjects were involved. 

We release \FreshBrew{} to support transparent and reproducible research in AI-assisted
software engineering. While the benchmark may help improve automated migration
systems, users should remain aware of potential misuse—such as over-reliance on
autonomous agents for code changes without human review. Responsible application
of these tools should always include developer oversight and verification of
functional correctness.

\clearpage
\bibliographystyle{unsrtnat}
\bibliography{references}  

@misc{liu2025migrationbenchrepositorylevelcodemigration,
      title={MigrationBench: Repository-Level Code Migration Benchmark from Java 8}, 
      author={Linbo Liu and Xinle Liu and Qiang Zhou and Lin Chen and Yihan Liu and Hoan Nguyen and Behrooz Omidvar-Tehrani and Xi Shen and Jun Huan and Omer Tripp and Anoop Deoras},
      year={2025},
      eprint={2505.09569},
      archivePrefix={arXiv},
      primaryClass={cs.SE},
      url={https://arxiv.org/abs/2505.09569}, 
}

@Misc{smolagents,
  title =        {`smolagents`: a smol library to build great agentic systems.},
  author =       {Aymeric Roucher and Albert Villanova del Moral and Thomas Wolf and Leandro von Werra and Erik Kaunismäki},
  howpublished = {\url{https://github.com/huggingface/smolagents}},
  year =         {2025}
}

@misc{wang2024executablecodeactionselicit,
      title={Executable Code Actions Elicit Better LLM Agents}, 
      author={Xingyao Wang and Yangyi Chen and Lifan Yuan and Yizhe Zhang and Yunzhu Li and Hao Peng and Heng Ji},
      year={2024},
      eprint={2402.01030},
      archivePrefix={arXiv},
      primaryClass={cs.CL},
      url={https://arxiv.org/abs/2402.01030}, 
}

@article{shyrobokov,
author = {Shyrobokov, Valentyn},
year = {2025},
month = {02},
pages = {},
title = {APPROACHES TO MIGRATING INFORMATION SYSTEMS TO MODERN JAVA VERSIONS},
volume = {131},
journal = {Universum:Technical sciences},
doi = {10.32743/UniTech.2025.131.2.19340}
}

@article{RAEMAEKERS2017140,
title = {Semantic versioning and impact of breaking changes in the Maven repository},
journal = {Journal of Systems and Software},
volume = {129},
pages = {140-158},
year = {2017},
issn = {0164-1212},
doi = {https://doi.org/10.1016/j.jss.2016.04.008},
url = {https://www.sciencedirect.com/science/article/pii/S0164121216300243},
author = {S. Raemaekers and A. {van Deursen} and J. Visser}
}

@misc{wang2025openhandsopenplatformai,
      title={OpenHands: An Open Platform for AI Software Developers as Generalist Agents}, 
      author={Xingyao Wang and Boxuan Li and Yufan Song and Frank F. Xu and Xiangru Tang and Mingchen Zhuge and Jiayi Pan and Yueqi Song and Bowen Li and Jaskirat Singh and Hoang H. Tran and Fuqiang Li and Ren Ma and Mingzhang Zheng and Bill Qian and Yanjun Shao and Niklas Muennighoff and Yizhe Zhang and Binyuan Hui and Junyang Lin and Robert Brennan and Hao Peng and Heng Ji and Graham Neubig},
      year={2025},
      eprint={2407.16741},
      archivePrefix={arXiv},
      primaryClass={cs.SE},
      url={https://arxiv.org/abs/2407.16741}, 
}

@misc{jimenez2024swebenchlanguagemodelsresolve,
      title={SWE-bench: Can Language Models Resolve Real-World GitHub Issues?}, 
      author={Carlos E. Jimenez and John Yang and Alexander Wettig and Shunyu Yao and Kexin Pei and Ofir Press and Karthik Narasimhan},
      year={2024},
      eprint={2310.06770},
      archivePrefix={arXiv},
      primaryClass={cs.CL},
      url={https://arxiv.org/abs/2310.06770}, 
}

@misc{misra2025gitchameleonevaluatingaicode,
      title={GitChameleon: Evaluating AI Code Generation Against Python Library Version Incompatibilities}, 
      author={Diganta Misra and Nizar Islah and Victor May and Brice Rauby and Zihan Wang and Justine Gehring and Antonio Orvieto and Muawiz Chaudhary and Eilif B. Muller and Irina Rish and Samira Ebrahimi Kahou and Massimo Caccia},
      year={2025},
      eprint={2507.12367},
      archivePrefix={arXiv},
      primaryClass={cs.SE},
      url={https://arxiv.org/abs/2507.12367}, 
}

@misc{recent-frontier-models-are-reward-hacking,
    title = {Recent Frontier Models Are Reward Hacking},
    author = {METR},
    howpublished = {\url{https://metr.org/blog/2025-06-05-recent-reward-hacking/}},
    year = {2025},
    month = {06},
  }

@misc{gpt4_1_openai_2025,
    author = {OpenAI},
    title = {{Introducing GPT-4.1 in the API}},
    howpublished = {\url{https://openai.com/index/gpt-4-1/}},
    year = {2025},
    month = {April},
    note = {Discusses GPT-4.1, GPT-4.1 mini, and GPT-4.1 nano}
}

@misc{zheng2024understandinglargelanguagemodels,
      title={Towards an Understanding of Large Language Models in Software Engineering Tasks}, 
      author={Zibin Zheng and Kaiwen Ning and Qingyuan Zhong and Jiachi Chen and Wenqing Chen and Lianghong Guo and Weicheng Wang and Yanlin Wang},
      year={2024},
      eprint={2308.11396},
      archivePrefix={arXiv},
      primaryClass={cs.SE},
      url={https://arxiv.org/abs/2308.11396}, 
}

@misc{hou2024largelanguagemodelssoftware,
      title={Large Language Models for Software Engineering: A Systematic Literature Review}, 
      author={Xinyi Hou and Yanjie Zhao and Yue Liu and Zhou Yang and Kailong Wang and Li Li and Xiapu Luo and David Lo and John Grundy and Haoyu Wang},
      year={2024},
      eprint={2308.10620},
      archivePrefix={arXiv},
      primaryClass={cs.SE},
      url={https://arxiv.org/abs/2308.10620}, 
}

@misc{he2024llmbasedmultiagentsystemssoftware,
      title={LLM-Based Multi-Agent Systems for Software Engineering: Literature Review, Vision and the Road Ahead}, 
      author={Junda He and Christoph Treude and David Lo},
      year={2024},
      eprint={2404.04834},
      archivePrefix={arXiv},
      primaryClass={cs.SE},
      url={https://arxiv.org/abs/2404.04834}, 
}

@misc{yang2024sweagentagentcomputerinterfacesenable,
      title={SWE-agent: Agent-Computer Interfaces Enable Automated Software Engineering}, 
      author={John Yang and Carlos E. Jimenez and Alexander Wettig and Kilian Lieret and Shunyu Yao and Karthik Narasimhan and Ofir Press},
      year={2024},
      eprint={2405.15793},
      archivePrefix={arXiv},
      primaryClass={cs.SE},
      url={https://arxiv.org/abs/2405.15793}, 
}

@misc{bairi2023codeplanrepositorylevelcodingusing,
      title={CodePlan: Repository-level Coding using LLMs and Planning}, 
      author={Ramakrishna Bairi and Atharv Sonwane and Aditya Kanade and Vageesh D C and Arun Iyer and Suresh Parthasarathy and Sriram Rajamani and B. Ashok and Shashank Shet},
      year={2023},
      eprint={2309.12499},
      archivePrefix={arXiv},
      primaryClass={cs.SE},
      url={https://arxiv.org/abs/2309.12499}, 
}

@misc{wang2025softwaredevelopmentlifecycle,
      title={Software Development Life Cycle Perspective: A Survey of Benchmarks for Code Large Language Models and Agents}, 
      author={Kaixin Wang and Tianlin Li and Xiaoyu Zhang and Chong Wang and Weisong Sun and Yang Liu and Bin Shi},
      year={2025},
      eprint={2505.05283},
      archivePrefix={arXiv},
      primaryClass={cs.SE},
      url={https://arxiv.org/abs/2505.05283}, 
}

@misc{chen2021evaluatinglargelanguagemodels,
      title={Evaluating Large Language Models Trained on Code}, 
      author={Mark Chen and Jerry Tworek and Heewoo Jun and Qiming Yuan and Henrique Ponde de Oliveira Pinto and Jared Kaplan and Harri Edwards and Yuri Burda and Nicholas Joseph and Greg Brockman and Alex Ray and Raul Puri and Gretchen Krueger and Michael Petrov and Heidy Khlaaf and Girish Sastry and Pamela Mishkin and Brooke Chan and Scott Gray and Nick Ryder and Mikhail Pavlov and Alethea Power and Lukasz Kaiser and Mohammad Bavarian and Clemens Winter and Philippe Tillet and Felipe Petroski Such and Dave Cummings and Matthias Plappert and Fotios Chantzis and Elizabeth Barnes and Ariel Herbert-Voss and William Hebgen Guss and Alex Nichol and Alex Paino and Nikolas Tezak and Jie Tang and Igor Babuschkin and Suchir Balaji and Shantanu Jain and William Saunders and Christopher Hesse and Andrew N. Carr and Jan Leike and Josh Achiam and Vedant Misra and Evan Morikawa and Alec Radford and Matthew Knight and Miles Brundage and Mira Murati and Katie Mayer and Peter Welinder and Bob McGrew and Dario Amodei and Sam McCandlish and Ilya Sutskever and Wojciech Zaremba},
      year={2021},
      eprint={2107.03374},
      archivePrefix={arXiv},
      primaryClass={cs.LG},
      url={https://arxiv.org/abs/2107.03374}, 
}

@misc{lu2021codexgluemachinelearningbenchmark,
      title={CodeXGLUE: A Machine Learning Benchmark Dataset for Code Understanding and Generation}, 
      author={Shuai Lu and Daya Guo and Shuo Ren and Junjie Huang and Alexey Svyatkovskiy and Ambrosio Blanco and Colin Clement and Dawn Drain and Daxin Jiang and Duyu Tang and Ge Li and Lidong Zhou and Linjun Shou and Long Zhou and Michele Tufano and Ming Gong and Ming Zhou and Nan Duan and Neel Sundaresan and Shao Kun Deng and Shengyu Fu and Shujie Liu},
      year={2021},
      eprint={2102.04664},
      archivePrefix={arXiv},
      primaryClass={cs.SE},
      url={https://arxiv.org/abs/2102.04664}, 
}

@misc{li2024evocodebenchevolvingcodegeneration,
      title={EvoCodeBench: An Evolving Code Generation Benchmark Aligned with Real-World Code Repositories}, 
      author={Jia Li and Ge Li and Xuanming Zhang and Yihong Dong and Zhi Jin},
      year={2024},
      eprint={2404.00599},
      archivePrefix={arXiv},
      primaryClass={cs.CL},
      url={https://arxiv.org/abs/2404.00599}, 
}

@misc{li2024devevalmanuallyannotatedcodegeneration,
      title={DevEval: A Manually-Annotated Code Generation Benchmark Aligned with Real-World Code Repositories}, 
      author={Jia Li and Ge Li and Yunfei Zhao and Yongmin Li and Huanyu Liu and Hao Zhu and Lecheng Wang and Kaibo Liu and Zheng Fang and Lanshen Wang and Jiazheng Ding and Xuanming Zhang and Yuqi Zhu and Yihong Dong and Zhi Jin and Binhua Li and Fei Huang and Yongbin Li},
      year={2024},
      eprint={2405.19856},
      archivePrefix={arXiv},
      primaryClass={cs.CL},
      url={https://arxiv.org/abs/2405.19856}, 
}

@misc{cassano2022multiplescalableextensibleapproach,
      title={MultiPL-E: A Scalable and Extensible Approach to Benchmarking Neural Code Generation}, 
      author={Federico Cassano and John Gouwar and Daniel Nguyen and Sydney Nguyen and Luna Phipps-Costin and Donald Pinckney and Ming-Ho Yee and Yangtian Zi and Carolyn Jane Anderson and Molly Q Feldman and Arjun Guha and Michael Greenberg and Abhinav Jangda},
      year={2022},
      eprint={2208.08227},
      archivePrefix={arXiv},
      primaryClass={cs.LG},
      url={https://arxiv.org/abs/2208.08227}, 
}

@misc{tao2024unravelingpotentiallargelanguage,
      title={Unraveling the Potential of Large Language Models in Code Translation: How Far Are We?}, 
      author={Qingxiao Tao and Tingrui Yu and Xiaodong Gu and Beijun Shen},
      year={2024},
      eprint={2410.09812},
      archivePrefix={arXiv},
      primaryClass={cs.SE},
      url={https://arxiv.org/abs/2410.09812}, 
}

@misc{liang2025rustevo2evolvingbenchmarkapi,
      title={RustEvo2: An Evolving Benchmark for API Evolution in LLM-based Rust Code Generation}, 
      author={Linxi Liang and Jing Gong and Mingwei Liu and Chong Wang and Guangsheng Ou and Yanlin Wang and Xin Peng and Zibin Zheng},
      year={2025},
      eprint={2503.16922},
      archivePrefix={arXiv},
      primaryClass={cs.SE},
      url={https://arxiv.org/abs/2503.16922}, 
}

@inproceedings{Zhang_2024, series={ICSE ’24},
   title={Learning-based Widget Matching for Migrating GUI Test Cases},
   url={http://dx.doi.org/10.1145/3597503.3623322},
   DOI={10.1145/3597503.3623322},
   booktitle={Proceedings of the IEEE/ACM 46th International Conference on Software Engineering},
   publisher={ACM},
   author={Zhang, Yakun and Zhang, Wenjie and Ran, Dezhi and Zhu, Qihao and Dou, Chengfeng and Hao, Dan and Xie, Tao and Zhang, Lu},
   year={2024},
   month=feb, pages={1–13},
   collection={ICSE ’24} }

@misc{austin2021programsynthesislargelanguage,
      title={Program Synthesis with Large Language Models}, 
      author={Jacob Austin and Augustus Odena and Maxwell Nye and Maarten Bosma and Henryk Michalewski and David Dohan and Ellen Jiang and Carrie Cai and Michael Terry and Quoc Le and Charles Sutton},
      year={2021},
      eprint={2108.07732},
      archivePrefix={arXiv},
      primaryClass={cs.PL},
      url={https://arxiv.org/abs/2108.07732}, 
}

@misc{comanici2025gemini25pushingfrontier,
      title={Gemini 2.5: Pushing the Frontier with Advanced Reasoning, Multimodality, Long Context, and Next Generation Agentic Capabilities}, 
      author={Gheorghe Comanici and Multiple Authors},
      year={2025},
      eprint={2507.06261},
      archivePrefix={arXiv},
      primaryClass={cs.CL},
      url={https://arxiv.org/abs/2507.06261}, 
}

@article{gpt4o_system_card_2024,
    author = {OpenAI},
    title = {{GPT-4o System Card}},
    journal = {arXiv preprint arXiv:2410.21276},
    year = {2024},
    url = {https://arxiv.org/abs/2410.21276},
    note = {Cited for GPT-4o}
}

@misc{openai_o3_system_card,
    author = {OpenAI},
    title = {{OpenAI o3-mini System Card}},
    howpublished = {\url{https://openai.com/index/o3-mini-system-card/}},
    year = {2025},
    note = {Discusses the o3-mini model}
}

@misc{deepseekai2025deepseekv3technicalreport,
      title={DeepSeek-V3 Technical Report}, 
      author={DeepSeek-AI and Aixin Liu and Multiple Authors},
      year={2025},
      eprint={2412.19437},
      archivePrefix={arXiv},
      primaryClass={cs.CL},
      url={https://arxiv.org/abs/2412.19437}, 
}

@misc{gu2025surveyllmasajudge,
      title={A Survey on LLM-as-a-Judge}, 
      author={Jiawei Gu and Xuhui Jiang and Zhichao Shi and Hexiang Tan and Xuehao Zhai and Chengjin Xu and Wei Li and Yinghan Shen and Shengjie Ma and Honghao Liu and Saizhuo Wang and Kun Zhang and Yuanzhuo Wang and Wen Gao and Lionel Ni and Jian Guo},
      year={2025},
      eprint={2411.15594},
      archivePrefix={arXiv},
      primaryClass={cs.CL},
      url={https://arxiv.org/abs/2411.15594}, 
}

@misc{yang2025qwen3technicalreport,
      title={Qwen3 Technical Report}, 
      author={An Yang and Anfeng Li and Baosong Yang and Beichen Zhang and Binyuan Hui and Bo Zheng and Bowen Yu and Chang Gao and Chengen Huang and Chenxu Lv and Chujie Zheng and Dayiheng Liu and Fan Zhou and Fei Huang and Feng Hu and Hao Ge and Haoran Wei and Huan Lin and Jialong Tang and Jian Yang and Jianhong Tu and Jianwei Zhang and Jianxin Yang and Jiaxi Yang and Jing Zhou and Jingren Zhou and Junyang Lin and Kai Dang and Keqin Bao and Kexin Yang and Le Yu and Lianghao Deng and Mei Li and Mingfeng Xue and Mingze Li and Pei Zhang and Peng Wang and Qin Zhu and Rui Men and Ruize Gao and Shixuan Liu and Shuang Luo and Tianhao Li and Tianyi Tang and Wenbiao Yin and Xingzhang Ren and Xinyu Wang and Xinyu Zhang and Xuancheng Ren and Yang Fan and Yang Su and Yichang Zhang and Yinger Zhang and Yu Wan and Yuqiong Liu and Zekun Wang and Zeyu Cui and Zhenru Zhang and Zhipeng Zhou and Zihan Qiu},
      year={2025},
      eprint={2505.09388},
      archivePrefix={arXiv},
      primaryClass={cs.CL},
      url={https://arxiv.org/abs/2505.09388}, 
}

@misc{zhang2024autocoderoverautonomousprogramimprovement,
      title={AutoCodeRover: Autonomous Program Improvement}, 
      author={Yuntong Zhang and Haifeng Ruan and Zhiyu Fan and Abhik Roychoudhury},
      year={2024},
      eprint={2404.05427},
      archivePrefix={arXiv},
      primaryClass={cs.SE},
      url={https://arxiv.org/abs/2404.05427}, 
}

@misc{zheng2025makeslargelanguagemodels,
      title={What Makes Large Language Models Reason in (Multi-Turn) Code Generation?}, 
      author={Kunhao Zheng and Juliette Decugis and Jonas Gehring and Taco Cohen and Benjamin Negrevergne and Gabriel Synnaeve},
      year={2025},
      eprint={2410.08105},
      archivePrefix={arXiv},
      primaryClass={cs.CL},
      url={https://arxiv.org/abs/2410.08105}, 
}

@misc{jain2025testgenevalrealworldunit,
      title={TestGenEval: A Real World Unit Test Generation and Test Completion Benchmark}, 
      author={Kush Jain and Gabriel Synnaeve and Baptiste Rozière},
      year={2025},
      eprint={2410.00752},
      archivePrefix={arXiv},
      primaryClass={cs.SE},
      url={https://arxiv.org/abs/2410.00752}, 
}

@misc{arceeModelSelection,
	author = {Arcee},
	title = {{M}odel {S}election | {A}rcee {A}{I} {D}ocumentation --- docs.arcee.ai},
	howpublished = {\url{https://docs.arcee.ai/arcee-conductor/arcee-small-language-models/model-selection##caller-large-tool-use-and-function-call}},
	year = {},
	note = {[Accessed 15-07-2025]},
}

@misc{Mockito,
        author = {Szczepan Faber and Brice Dutheil, Rafael Winterhalter and Tim van der Lippe et al.},
	title = {Mockito framework site},
	url = {https://site.mockito.org/},
	urldate = {2025-07-18},
}

@misc{SLF4J,
  author = {{QOS-ch}},
  title = {{SLF4J: Simple Logging Facade for Java}},
  howpublished = {\url{https://www.slf4j.org/}},
  year = {2025}
}

@misc{JacksonDatabind,
  author = {{FasterXML}},
  title = {{Jackson-databind: General data-binding for Jackson}},
  howpublished = "\url{https://github.com/FasterXML/jackson-databind}",
  year = {2024}
}

@misc{JEP260,
  author = {Mark Reinhold},
  title = {{JEP 260: Encapsulate Most Internal APIs}},
  howpublished = {OpenJDK JEP},
  year = {2015},
  url = {https://openjdk.org/jeps/260}
}

@misc{JEP403,
  author = {Alex Buckley and Mark Reinhold},
  title = {{JEP 403: Strongly Encapsulate JDK Internals}},
  howpublished = {OpenJDK JEP},
  year = {2021},
  url = {https://openjdk.org/jeps/403}
}

@misc{OracleJDKMigrationGuide,
  author = {{Oracle}},
  title = {{Oracle JDK Migration Guide: Migrating from JDK 8 to Later JDK Releases}},
  howpublished = {Oracle Java Documentation},
  year = {2026},
  url = {https://docs.oracle.com/en/java/javase/26/migrate/migrating-from-jdk-8-later-jdk-releases.html}
}

@misc{JEP471,
  author = {Ron Pressler and Alex Buckley},
  title = {{JEP 471: Deprecate the Memory-Access Methods in sun.misc.Unsafe for Removal}},
  howpublished = {OpenJDK JEP},
  year = {2024},
  url = {https://openjdk.org/jeps/471}
}
\newpage

\appendix

\section{Benchmark - Additional Details}
\begin{figure}[!htbp]
    \centering
    \includegraphics[width=0.5\linewidth]{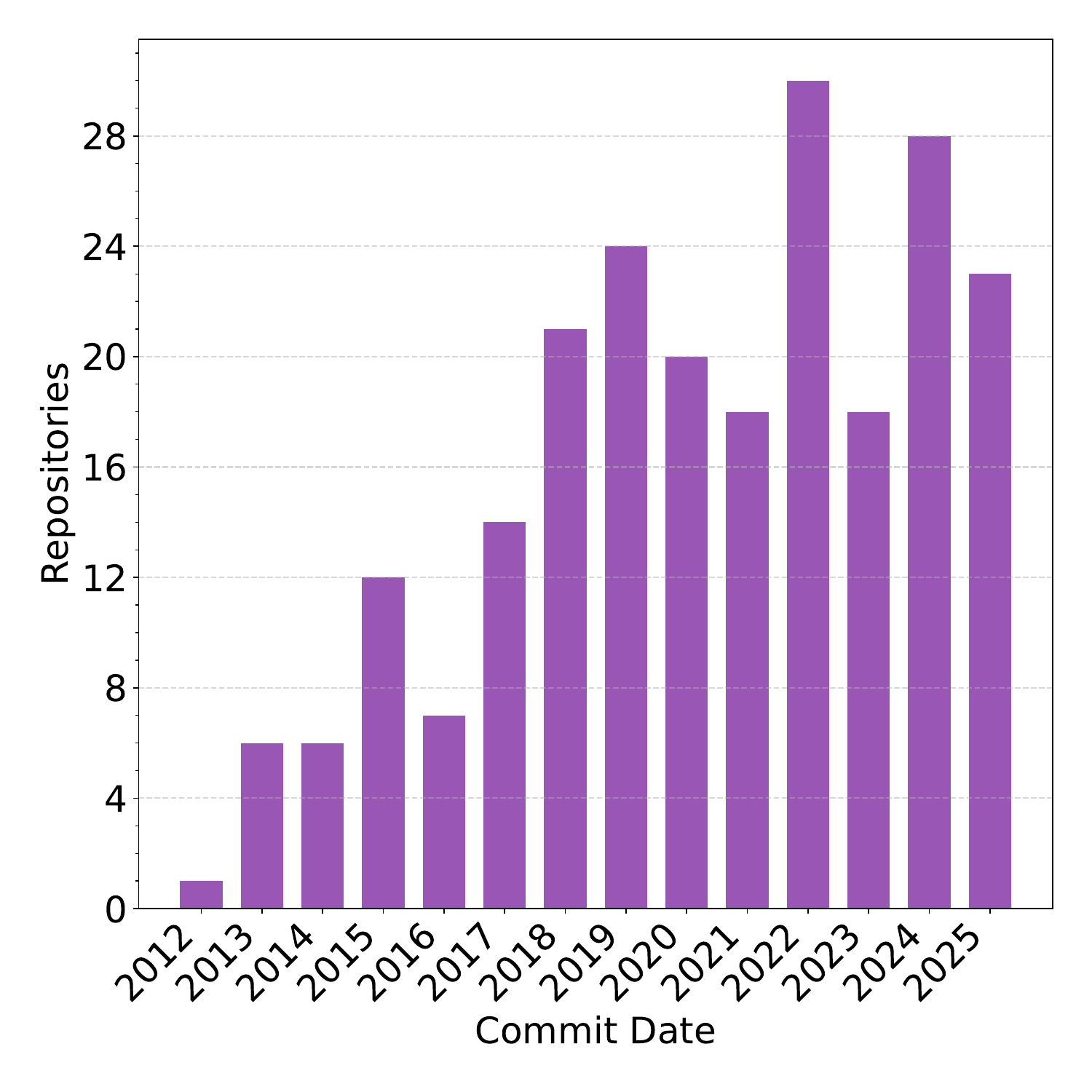} \caption{The temporal distribution of the dataset, showing the number of repositories based on the year of their sampled commit. The dataset is primarily composed of modern projects, with a high concentration from 2018 onwards.}
    \label{fig:commit_dates}
\end{figure}

Figure~\ref{fig:commit_dates} presents details on the temporal distribution of the dataset.

\paragraph{Library/Application Composition.} \FreshBrew{} currently reports structural statistics such as dependencies, modules, tests, lines of code, and coverage. We do not yet classify repositories as libraries, frameworks, plugins, tools, or applications. Such a classification would help quantify potential dataset skew and support more fine-grained reporting of migration difficulty by project type. We leave this characterization to future versions of the benchmark.

\section{Prompts}
This appendix contains the full text of the prompts used in our experiments to ensure reproducibility.

Figure~\ref{fig:agent_prompt_template} shows the primary instruction given to the AI agent for the migration task. The prompt clearly defines the objective (upgrade to a target JDK version), the success criteria (build and tests must pass), and includes a critical constraint forbidding the agent from deleting or commenting out tests. This instruction is central to our study's focus on preventing reward hacking by guiding the agent toward genuine problem-solving.

Figure~\ref{fig:failure_mode_prompt_template} presents the prompt given to the LLM-as-judge for our qualitative failure analysis. This prompt establishes the judge's persona as an expert, provides a strict classification schema for failure modes, and specifies a structured JSON output to ensure consistent and parsable results from the analysis.

\begin{figure}[htbp]
  \centering
  
\begin{promptcode}
  Upgrade the Maven project rooted at the current working dir to use JDK {version}. Ensure that the build and the tests pass. You can make source code and dependency changes. You cannot delete or comment-out tests. To do a final verification of your work before calling the `final_answer` tool, verify your work by making sure that `maven_verify` tool succeeds with the parameter `clean` set to `True`, because this is how we are going to evaluate your success. DO NOT call maven_verify with `clean=True` before ensuring it passes without it, because its slow and outputs a lot of tokens.
    \end{promptcode}

\caption{The prompt template used for our experiments with \textit{smolagents}.}
\label{fig:agent_prompt_template}
\end{figure}

\begin{figure}[htbp]
  \centering
  
\begin{promptcode}
  You are an expert Java software engineer and researcher specializing in code migration and developer tool evaluation. Your task is to analyze the final {{n_steps}} 'thought-action' steps from a failed attempt by an AI agent to migrate a Java 8 project to Java 17.
Based on the provided trace, identify the primary technical reason for the failure. Do not simply state that the agent failed or ran out of steps. Pinpoint the specific build, dependency, or code-level issue that the agent was unable to resolve.
Choose ONLY ONE of the following categories that best describes the failure:
* Dependency Management Failure
* Build Configuration Error
* Java API Incompatibility
* Agent Behavioral Failure
* Root Cause Not in Final Steps
* Unknown
The agent's final {{n_steps}} steps are as follows:
--- BEGIN TRACE ---
{{final_steps_trace}}
--- END TRACE ---
Provide your output in JSON format with two keys: "failure_category" and "reasoning". The reasoning should be a brief, one-sentence explanation supporting your choice.
    \end{promptcode}

\caption{The prompt used for failure mode analysis.}
\label{fig:failure_mode_prompt_template}
\end{figure}

\clearpage

\end{document}